\documentclass[aps,superscriptaddress,tightenlines,nofootinbib,floatfix,longbibliography,notitlepage]{revtex4-1}
\usepackage[left=18mm,right=19mm,top=23mm,bottom=16mm]{geometry}

\usepackage[T1]{fontenc}
\usepackage[utf8x]{inputenc}
\usepackage[english]{babel}
\usepackage{bbold}
\usepackage{amsmath,amssymb}
\usepackage{bm}
\usepackage{comment}
\usepackage{graphicx}
\usepackage[dvipsnames]{xcolor}
\usepackage{slashed}
\usepackage{stmaryrd}
\usepackage[
    colorlinks=true,
    allcolors=blue
]{hyperref}

\providecommand{\repositoryInformationSetup}{} 
\repositoryInformationSetup

\usepackage{xspace}
\usepackage{dsfont}

\newcommand{\repoURL}{https://github.com/evanberkowitz/latex-base}

\newcommand{\secref}[1]{Sec.~\ref{sec:#1}}
\newcommand{\Secref}[1]{Section~\ref{sec:#1}}

\newcommand{\Appref}[1]{Appendix~\ref{sec:#1}}

\newcommand{\Figref}[1]{Figure~\ref{fig:#1}\xspace}
\renewcommand{\eqref}[1]{(\ref{eq:#1})\xspace}

\def\Ref#1{Ref.~\cite{#1}}  

\newcommand{\Refs}[1]{Refs.~\cite{#1}}

\newcommand{\goesto}{\ensuremath{\rightarrow}}

\newcommand{\one}{\ensuremath{\mathds{1}}}

\newcommand{\Reals}{\ensuremath{\mathds{R}}\xspace}
\newcommand{\Complexes}{\ensuremath{\mathds{C}}\xspace}

\newcommand{\pp}{{{}^{+}_+}}
\newcommand{\mm}{{{}^{-}_-}}

\newcommand{\oneover}[1]{\ensuremath{\frac{1}{#1}}}                             
\newcommand{\inverse}{\ensuremath{^{-1}}}                                       
\newcommand{\half}{\ensuremath{\frac{1}{2}} }                                   

\newcommand{\dif}{\ensuremath{\text{d}}}
\newcommand{\Dif}{\ensuremath{\mathcal{D}}}
\newcommand{\od}[2]{\ensuremath{\frac{\text{d}#1}{\text{d}#2}}}
\newcommand{\pd}[2]{\ensuremath{\frac{\partial#1}{\partial#2}}}

\newcommand{\abs}[1]{\ensuremath{\left| #1 \right|}\xspace}

\newcommand{\tr}[1]{\ensuremath{\text{tr}\left[#1\right]}}

\newcommand{\adjoint}[1]{\ensuremath{{#1}^{\dagger}}}

\newcommand{\C}[1]{\ensuremath{C_{#1}}\xspace}
\renewcommand{\O}{\ensuremath{\hat{O}}\xspace}
\newcommand{\Z}{\ensuremath{\mathcal{Z}}\xspace}
\newcommand{\Sr}{\ensuremath{S^R}\xspace}
\newcommand{\Si}{\ensuremath{S^I}\xspace}

\newcommand{\Nt}{\ensuremath{N_t}\xspace}

\let\builtinLaTeX\LaTeX
\def\LaTeX{\builtinLaTeX\xspace}

\newcommand{\Seff}{\ensuremath{S_{\text{eff}}}}
\renewcommand{\Re}{\text{Re}}
\renewcommand{\Im}{\text{Im}}
\newcommand{\hash}{\texttt{\#}\xspace}
\newcommand{\Utilde}{\ensuremath{\tilde{U}}}

\newcommand{\nt}{\ensuremath{N_t}\xspace}
\newcommand{\NN}{\ensuremath{\text{NN}}\xspace}

\newcommand{\umd}{
    Maryland Center for Fundamental Physics,
    University of Maryland, College Park, MD 20742, USA
}

\newcommand{\ikp}{
    Institut f\"{u}r Kernphysik,
    Forschungszentrum J\"{u}lich, 54245 J\"{u}lich Germany
}

\newcommand{\ias}{
    Institute for Advanced Simulation,
    Forschungszentrum J\"{u}lich, 54245 J\"{u}lich Germany
}

\newcommand{\bonn}{
    Helmholtz-Institut f\"{u}r Strahlen- und Kernphysik,
    Rheinische Friedrich-Wilhelms-Universit\"{a}t Bonn, 53012 Bonn Germany
}

\newcommand{\jsc}{
    JARA-HPC, J\"{u}lich Supercomputing Center,
    Forschungszentrum J\"{u}lich, 54245 J\"{u}lich Germany
}

\begin{document}

\title{Leveraging Machine Learning to Alleviate Hubbard Model Sign Problems}

\author{Jan-Lukas Wynen}        \affiliation{\ias}
\author{Evan Berkowitz}         \affiliation{\umd} \affiliation{\ias}
\author{Stefan Krieg}           \affiliation{\ias} \affiliation{\jsc}
\author{Thomas Luu}             \affiliation{\ias} \affiliation{\ikp} \affiliation{\bonn}
\author{Johann Ostmeyer}        \affiliation{\bonn}

\date{\today}

\begin{abstract}
Lattice Monte Carlo calculations of interacting systems on non-bipartite lattices exhibit an oscillatory imaginary phase known as the phase or sign problem, even at zero chemical potential.
One method to alleviate the sign problem is to analytically continue the integration region of the state variables into the complex plane via holomorphic flow equations.
For asymptotically large flow times the state variables approach manifolds of constant imaginary phase known as Lefschetz thimbles.
However, flowing such variables and calculating the ensuing Jacobian is a computationally demanding procedure.
In this paper we demonstrate that neural networks can be trained to parameterize suitable manifolds for this class of sign problem and drastically reduce the computational cost.
We apply our method to the Hubbard model on the triangle and tetrahedron, both of which are non-bipartite.
At strong interaction strengths and modest temperatures the tetrahedron suffers from a severe sign problem that cannot be overcome with standard reweighting techniques, while it quickly yields to our method.
We benchmark our results with exact calculations and comment on future directions of this work.
\end{abstract}

\maketitle

\section{Introduction}\label{sec:intro}

Lattice field theories allow for a first-principles construction of non-perturbative interacting quantum field theories.
Beyond being a mathematical footing, they provide a computational strategy for solving such systems numerically, typically via Markov-chain Monte Carlo (MCMC).
However, for this numerical approach to succeed, a Euclidean field theory requires a real-valued action, providing a positive-definite integration measure.
When the action is complex, additional steps must be taken, because the integrand can oscillate wildly. Hence, intricate cancellations are required for numerical estimates to yield accurate results, thereby rendering otherwise successful lattice methods powerless.

This oscillating complex phase problem (``sign problem'') is prevalent in many areas of computational physics that rely on lattice stochastic methods to tackle non-perturbative phenomena.

Lattice quantum chromodynamics (LQCD) calculations at finite baryon chemical potential~\cite{Splittorff:2007ck,Danzer:2009dk} suffer from a sign problem, which precludes any numerical investigation of quark matter in dense astrophysical objects such as neutron stars and supernovae~\cite{Garron:2016noc,Hsu:2010zza,Goy:2016egl}.
Furthermore, including the strong-$\theta$ term~\cite{Shindler:2015aqa} in the QCD action induces a sign problem as well.
Without the development of methods to alleviate the sign problem, one is forced to assume the affected terms are (perturbatively) small ~\cite{Splittorff:2007ck,Gavai:2008zr,Dragos:2019oxn}.

In nuclear lattice effective field theory (NLEFT)~\cite{Lahde:2019npb,Lahde:2015ona,Elhatisari:2017eno}, where nucleons are degrees of freedom as opposed to quarks, the sign problem prevents lattice MCMC studies of very neutron-rich nuclei.
These nuclei, near the neutron drip line, play an important role in today's nuclear reactors and in the industrial and astrophysical synthesis of heavy elements.

A wide variety of condensed-matter systems, including the doped fermionic Hubbard model, exhibit a rich multi-quasi-particle spectrum (see, for example, \Refs{PhysRevLett.106.037404,doi:10.1021/acs.nanolett.7b03111,PhysRevLett.123.167401,doi:10.1021/acsnano.9b07207}) and comprise tantalizing theoretical systems with commercial relevance. Unfortunately, many of these systems exhibit a sign problem as well. Indeed, the sign problem poses a major stumbling block for MCMC studies in all computational subdisciplines of physics.

Finding a general solution to the phase problem with polynomial scaling in the severity of the problem is NP-hard~\cite{Troyer:2004ge}.
However, techniques that take advantage of a particular model's structure may still be achievable.
Therefore, various strategies for alleviating the sign problem have been developed.

The most obvious and widespread of these is \emph{reweighting}, which we describe in more detail \Secref{reweighting}.
Reweighting can be applied when the sign problem is mild, but can fail spectacularly when the problem is severe, much as perturbation theory fails when the interactions become strong.

Another commonly used strategy is analytic continuation.
For example, in QCD one may simulate with purely imaginary baryon chemical potential, removing the complex phase completely (see, for example, \Refs{DElia:2002tig,Bellwied:2015rza,Vovchenko:2017xad}).  
Here, the uncertainty lies in the functional form chosen to analytically continue results back to the real axis, which is not known \emph{a priori} and thus relies on model assumptions.
Moreover, analytically continuing Monte Carlo data with uncertainties is no easy task.

Alternatively, Complex Langevin methods do away completely with MCMC and have had various degrees of success (see, for example, \Refs{Batrouni:1985jn,Batrouni:1985ye,Fukugita:1986tg,DAMGAARD1987227}, and more recently, \cite{Kogut:2019qmi}).
Unfortunately, there currently seems to be no consensus on which systems have Langevin methods that are guaranteed to be correct and which do not, though progress has been made in quantifying certain conditions for success \cite{Sexty:2014zya}.

Recently, tensor networks~\cite{DMRG_original,verstraete2004renormalization,Orus:2013kga} have shown promise in tackling many-body systems in both one and two dimensions.
Their formalism is agnostic to the presence of a chemical potential.
First results are available for fermionic systems~\cite{tensor_fermions_derivation}, such as spinless fermions on the hexagonal lattice~\cite{iPEPS_hubbard_2018} and the Hubbard model on a square lattice~\cite{iPEPS_hubbard_2016}.
It remains to be seen, however, whether such calculations are preferable in terms of precision, scalability, and computational complexity, which also applies to methods that involve direct integration over the group manifold providing polynomial exactness~\cite{Ammon:2016jap}.

The method we leverage here is related to integration on Lefschetz thimbles.
By analytically continuing the integration variables into the complex plane~\cite{Cristoforetti:2012su}, one can locate higher-dimensional steepest-descent analogues called \emph{Lefschetz thimbles} for each critical point\footnote{In this context, at a critical point is defined as a point in field space where the derivative of the action w.r.t. the (complex) field vanishes, see \eqref{critpflow}.}.
On a thimble the phase is not oscillatory, but constant, up to the residual phase of the Jacobian (which can reintroduce wild oscillations if the thimble is strongly curved~\cite{Lawrence:2018mve}).
Each thimble has its own constant phase; lattice Monte Carlo on a given thimble can be performed because the phase is global and can be factored out of the integral.
A combination of all thimbles which can be reached by \emph{holomorphic flow} yields the original integral over the real variables, but finding those thimbles (or equivalently, identifying their critical points) is usually difficult.

Early investigations with this method looked at bosonic gauge systems in low dimensions~\cite{Cristoforetti:2014gsa,Cristoforetti:2013wha,Mukherjee:2013aga}.
The Thirring model --- non-relativistic fermions with a chemical potential in one dimension --- was studied using only one ``main'' thimble~\cite{Fujii:2015bua,Fujii:2015vha,Kanazawa:2014qma} before yielding to a flow-based method~\cite{Alexandru:2015sua,Alexandru:2015xva,Tanizaki:2016lta} that approaches a thimble in the limit of long flow times.
Moreover, Lefschetz thimbles have found use in higher dimensions, including the 1+1-dimensional Thirring model~\cite{Alexandru:2016ejd} and small examples of the doped 2+1-dimensional Hubbard model~\cite{Ulybyshev:2019hfm}.
Gauge symmetry complicates the story but the efficacy of Lefschetz Thimbles in gauge theories is a field of active research~\cite{Schmidt:2017gvu,Zambello:2018ibq,Alexandru:2018ngw,Pawlowski:2020kok}.
Beyond Lefschetz thimbles one may find sign-optimized manifolds where even the residual Jacobian phase is handled cleanly~\cite{Alexandru:2018fqp,Alexandru:2018ddf,Mori:2017nwj}.

The most difficult aspect of attempting a Lefschetz decomposition is that the thimbles' locations and shapes are not generally known \emph{a priori}, and that their determination is a complicated and numerically intensive endeavor, especially as systems get larger in higher dimensions.
Even when locating critical points is straightforward, deciding whether their thimbles must be included to reproduce the integral of interest is not.

Numerical difficulties abound, as well.
For example, the calculation of the Jacobian associated with the transformation from the real plane to these complex manifolds and its determinant can be numerically prohibitive as the system size becomes larger.
These issues make an exact Lefschetz Thimble decomposition potentially as difficult as the original sign problem.

Instead one can approximate the thimbles by use of holomorphic flow equations.
Rather than completely solving the sign problem, this strategy may merely alleviate it~\cite{Alexandru:2015sua,Nishimura:2017vav}.
Moreover, numerical techniques to estimate the determinant of the Jacobian in an efficient manner fix scaling problems~\cite{Alexandru:2016lsn}.
Still, the determination of these approximate manifolds is less computationally intensive than determining the exact manifolds, especially since they live in high-dimensional spaces.

Machine learning provides a good tool for parameterizing the flowed manifold~\cite{Alexandru:2017czx}.
We can train a neural network to parameterize an approximate flowed manifold because these manifolds are continuous and smooth.
Then, manifolds defined via networks, or \emph{learnifolds}, can be integrated on to alleviate the sign problem.
The gain in speed by use of neural networks can be substantial.

We study the Hubbard model on small non-bipartite lattices which suffer a severe phase problem by incorporating a learnifold into HMC.\@
Even though standard reweighting techniques are completely ineffective, even for some of these small problems, we find that learnifold-HMC allows us to extract correlation functions well, reproducing exact results.
In the future we plan to leverage this technique to study fullerenes, such as buckyballs, and anticipate a straightforward application to doped systems as well.

Our paper is organized as follows.
In \Secref{formalism} we describe our formalism in detail.
In \Secref{algorithm} we then describe our algorithm, giving details about how we incorporate machine learning into HMC without compromising the algorithm's exactness.
In \Secref{results} we leverage our method, and show that it reproduces exact results for a number of different observables on some simple systems that may be solved exactly, even when standard reweighting might fail.
We point out that there is generally a trade-off between ergodicity and the sign problem.
Finally, we give some conclusions in \Secref{conclusions}.


\section{Formalism}\label{sec:formalism}
In this section we introduce the Hamiltonian used in our studies and discuss the consequences of having a non-bipartite lattice.
We then explain \emph{reweighting}---an exact method for handling complex actions.
Finally, we discuss correlation functions and mention the operators we use to construct them.

\subsection{The Hamiltonian}
\label{sec:hamiltonian}

We use the Hubbard model in the particle/hole basis~\cite{Brower:2012zd, Ulybyshev:2013swa, Smith:2014tha, Luu:2015gpl, Wynen:2018ryx} to perform simulations.
The Hubbard model consists of a tight-binding Hamiltonian,
\begin{equation}\label{eq:HO}
    H_{0}
    =
    - \sum_{x, y}\left(
            a_{x, \uparrow}^{\dagger} h_{xy}a_{y, \uparrow}
        +   a_{x, \downarrow}^{\dagger} h_{xy} a_{y, \downarrow}
        \right)\ ,
\end{equation}
where $h_{xy}$ is nonzero if $x$ and $y$ are nearest neighbors,
coupled with an onsite interaction of the form
\begin{equation}\label{eq:hubbard 1}
    H
    =
    H_{0}
    -\frac{U}{2} \sum_{x}\left(n_{x, \uparrow}-n_{x, \downarrow}\right)^{2}\ ,
\end{equation}
where the number operator $n_{x, s} \equiv a_{x,s}^\dag a^{}_{x,s}$ counts electrons of spin $s$ at position $x$.  We now change to the hole basis for the spin-$\downarrow$ electrons,
\begin{equation}\label{eq:particle-hole transformation}
b_{x, \downarrow}^{\dagger} \equiv a_{x, \downarrow}, \quad b_{x, \downarrow} \equiv a_{x, \downarrow}^{\dagger}\ ,
\end{equation}
which gives, up to an irrelevant constant,
\begin{align}\label{eq:hubbard 2}
    H
    &=
    -   \sum_{x, y}\left(
            a_{x}^{\dagger} h_{xy} a_{y}
        -   b_{x}^{\dagger} h_{xy} b_{y}
        \right)
    +   \frac{U}{2} \sum_{x}\rho_x^{2}\ ,
    \\
    \label{eq:charge-density}
    \rho_x
    &=
    n^a_{x}-n^b_{x}
\end{align}
where $n^a_x=a_x^\dagger a_x$ counts the number of (spin-$\uparrow$) particles $n^b_{x}=b_{x}^{\dag} b^{}_{x}$ counts the number of spin-$\downarrow$ holes at site $x$; we use the convention of positively-charged particles.

\subsection{Non-bipartite lattices}
\label{sec:non-bipartite}

Bipartite graphs are those that admit a two-coloring, such that no vertex has a neighbor of the same color.
Examples include the standard square lattice (consisting of two underlying square lattices), the honeycomb lattice (consisting of two underlying triangular lattices), or simply two connected sites.
A non-bipartite graph, in contrast, cannot be so colored.

Graphs with odd-length cycles are not bipartite.
Examples include fullerene refinements of a 2-sphere, such as the buckyball (C$_{60}$) or the dodecahedron (C$_{20}$).
In this case the presence of 12 pentagonal faces (required to make the geometry closed) destroys the bipartiteness.
The simplest, non-trivial non-bipartite graph is a single triangle, which is small enough for exact diagonalization.
The tetrahedron, topologically the complete graph on four vertices, is similarly tractable but not bipartite.
The Hubbard model has been studied on non-bipartite quasi-one-dimensional chains~\cite{PhysRevB.62.8658,PhysRevB.72.125116,PhysRevE.85.061123} and has been exactly solved on small clusters~\cite{doi:10.1002/andp.200710281} because solutions on such chains and clusters may be taken as input data for many-body methods.
The discovery of unusual heavy-fermion behavior of LiV${}_2$O${}_4$ triggered direct studies of the Hubbard model on the pyrochlore lattice with tetrahedral unit cells~\cite{PhysRevB.79.035115} and other extended non-bipartite structures~(for example, see~\cite{doi:10.1143/JPSJ.77.104702,PhysRevB.78.165113,PhysRevB.82.161101}).

As discussed in \Ref{Wynen:2018ryx}, when the Hubbard model is formulated on a bipartite graph, its formulation as a lattice field theory exhibits special features.
Discretizing the Euclidean time $\beta$ into \nt timeslices yields a temporal lattice spacing
\begin{equation}\label{eq:delta}
    \delta = \frac{\beta}{\nt}
\end{equation}
and we denote quantities made dimensionless with factors of $\delta$ with a tilde, so that $\Utilde=U\delta$.
The partition function can be cast into the form of a path integral~\cite{Brower:2012zd, Ulybyshev:2013swa, Smith:2014tha, Luu:2015gpl} yielding
\begin{align}
  \mathcal{Z} &=  \int\left[\prod_{x, t} \dif \phi_{x t}\right]e^{-\beta H[\phi]}
     =  \int\left[\prod_{x, t} \dif \phi_{x t}\right]{W[\phi]}
    \\
  W[\phi] &=
        \det \left(M^{}[\phi,h] \adjoint{M}[\phi,-h]\right)
        \exp \left(-\frac{1}{2 \tilde{U}} \sum_{x, t} \phi_{x t}^{2}\right)\ ,
        \label{eq:weight}
\end{align}
where we assume $\tilde{U} > 0$ and the fermion matrix is in the \emph{exponential discretization} (see \Ref{Wynen:2018ryx} for a comparison with other discretizations),
\begin{align}
  {M[\phi,h]}_{x^\prime t^\prime, xt}
  =
        \delta_{x^\prime, x} \delta_{t^\prime, t}
    -   {[e^{\tilde{h}}]}_{x^\prime, x} e^{i\phi_{xt}} \mathcal{B}_{t'} \delta_{t^\prime, t + 1},
\end{align}
where $\tilde{h}$ is the dimensionless hopping matrix and $\mathcal{B}_{t} = +1$ for $0 < t < N_t$ and $\mathcal{B}_0 = -1$ explicitly encodes anti-periodic temporal boundary conditions.

In the case of bipartite lattices, the particle-hole transformation~\eqref{particle-hole transformation} can be modified to include an additional sign
\begin{align}
  b_{x, \downarrow}^{\dagger}   &\equiv        \mathcal{P}_{x} a_{x, \downarrow}
  &
  b_{x, \downarrow}             &\equiv         \mathcal{P}_{x} a_{x, \downarrow}^{\dagger}\ ,
  \label{eq:particle hole with parity}
\end{align}
where $\mathcal{P}_x$ is the parity of the sublattice, so that we perform a site-dependent sign flip---$\mathcal{P}_x=+1$ if $x$ is on one sub-lattice, $-1$ if on the other---for holes.
This flips the sign of the hopping term in the hole matrix such that the weight becomes
\begin{align}
    W[\phi]
    &=
    \det \left(M^{}[\phi,h] \adjoint{M}[\phi,h]\right)
    \exp \left(-\frac{1}{2 \Utilde} \sum_{x, t} \phi_{x t}^{2}\right)
\end{align}
as shown in, for example, \Refs{Brower:2012zd, Ulybyshev:2013swa, Smith:2014tha, Luu:2015gpl}.
Since $M\adjoint{M}$ is positive-semidefinite, $W$ is real and positive-semi-definite as well\footnote{In the case of non-zero chemical potential, $W$ is complex-valued even for bipartite lattices. This has been investigated recently for small bipartite systems in the context of holomorphic flow~\cite{Ulybyshev:2019fte,Ulybyshev:2019hfm}}.
Such systems are thus easily amenable to standard Monte-Carlo simulations.

On non-bipartite lattices the signed particle-hole transformation~\eqref{particle hole with parity} does not exist, as the bipartitioning fails, so we cannot apply it and the weight \eqref{weight} is of indefinite sign.
The next section describes how Monte-Carlo techniques can nonetheless be applied in this case and discusses the problems such calculations typically face.

We restrict our attention to cases where the hopping matrix $h$ is always just a constant times the graph's adjacency matrix,
\begin{equation}
    h_{xy} = \kappa\; \delta_{\langle x,y \rangle}
\end{equation}
though this assumption could be relaxed to model realistic molecules where bond lengths and corresponding hopping strengths vary.
In general, the symmetry of the underlying lattice should be considered when determining the Hamiltonian.
For example, the regular dodecahedron is vertex- and edge- transitive, so it is natural to assign uniform interaction strengths to every site and uniform hopping strengths along every edge.
In contrast, while the truncated icosahedron (\C{60} buckyball) is vertex-transitive and thus every site should have the same interaction strength, it is not edge-transitive: some edges separate two hexagons, while others separate a hexagon from a pentagon and these two different kinds of edges could have different hopping strengths (which physically reflects the fact that the bond lengths differ).
In larger fullerenes with $I_h$ symmetry there are a wider variety of bond lengths, even between hexagonal faces, because some hexagons are closer to or farther from pentagons; in fullerenes with smaller symmetry groups (such as \C{70}, which enjoys a $D_{5h}$ symmetry) we can naturally incorporate its structure by adjusting hopping strengths in $h$ and adjusting $U$ from site to site in a way that respects its symmetry.
Since we here are interested in proof-of-principle work we ignore all effects of this kind.


\subsection{Reweighting}\label{sec:reweighting}

The expectation value of an observable $\O$ is given by
\begin{equation}
  \langle \O \rangle = \frac{1}{\Z} \int \Dif\phi\, \O[\phi]\ e^{-S[\phi]}
  \label{eq:expectval}
\end{equation}
where $S$ is the action, $\phi$ the fields, and $\Z$ the integral without the operator.
For the purpose of this discussion we include the fermionic determinants in the action $S$ by taking the log and casting them up into the exponential.

When the action is real-valued we can construct a Monte Carlo method, sampling configurations of the field $\phi$ according to their probability given by the Boltzmann weight $\exp(-S)/\Z$ (``importance sampling'').
Generated in such a way, we can estimate the expectation value in \eqref{expectval} by computing the mean of the observable measured separately on each configuration.

When the action is complex-valued the Boltzmann weight is complex as well, and does not directly provide a probability distribution.
\emph{Reweighting} is an exact, straightforward procedure by which we can overcome this difficulty, given sufficient computational resources.
Rather than sampling according to the action, one samples according to the real part of the action \Sr and incorporates the phase associated with the imaginary part of the action \Si into each observable, estimating
\begin{equation}\label{eq:reweighting}
  \langle \O \rangle = \frac{\langle\O e^{-i \Si}\rangle_R}{\langle e^{-i \Si}\rangle_R} \approx \frac{\sum_j \O\left[\phi_j\right] e^{-i\Si\left[\phi_j\right]}}{\sum_j e^{-i\Si\left[\phi_j\right]}}
\end{equation}
where $j$ runs over the ensemble and the $R$-subscripted angle brackets indicate an expectation value with respect to the real part of the action only\footnote{Unless otherwise mentioned, uncertainties presented here via reweighting come from a correlated bootstrap procedure, where, on each bootstrap resampling, the numerator and denominator are both measured and divided.}.

If the phase given by the imaginary part of the action is constant or narrowly distributed, this procedure can successfully estimate observables.
However, if the phase is widely distributed or, in the worst case, evenly covers the unit circle, the expectation value in the denominator nears zero and reweighting becomes computationally intractable.
We call the absolute value of the denominator the \emph{statistical power}
\begin{align}
  \Sigma = \abs{\langle e^{i\theta} \rangle_R}\ , \qquad \theta \equiv \arg W\ .
\end{align}
When all configurations have the same imaginary action, $\Sigma = 1$ and each configuration is valuable.
When $\theta[\phi]$ varies strongly for different field configurations $\phi$, $\Sigma$ is near zero and the configurations tend to cancel; each additional configuration contributes only marginally to the expectation value.
\emph{Quenching} the phase, that is, estimating expectation values by the uncontrolled approximation $\langle \O \rangle \simeq \langle \O \rangle_R$, can lead to a dramatic distortion of observables.


\subsection{Correlation Functions}\label{sec:correlators}

Two-point correlation functions between two operators $\O_x$ and $\adjoint{\O}_y$ at different times and sites $x$, $y$
\begin{equation}
    C_{xy}(\tau)
    = \oneover{\Z}\tr{
        e^{-\beta H}
        \oneover{N_t}\sum_t
        \O_x(t+\tau) \adjoint{\O}_y(t)
    }
    =
    \oneover{N_t}\sum_t
    \left\langle
        \O_x(t+\tau) \adjoint{\O}_y(t)
    \right\rangle
\end{equation}
can be computed as a function of temporal separation by solving for propagators on each configuration and tying them together in the required Wick contractions.
They admit spectral decompositions
\begin{align}
    C_{xy}(\tau)
    &= \oneover{\Z}\tr{
        e^{-\beta H}
        \oneover{N_t}\sum_t
        \O_x(t+\tau) \adjoint{\O}_y(t)
    }
    =
    \oneover{N_t}\sum_t
    \oneover{\Z}\tr{
        e^{-\beta H}\
        e^{+H(t+\tau)}\O_xe^{-H(t+\tau)}\
        e^{+Ht}\adjoint{\O}_ye^{-Ht}
    }
    \nonumber\\
    &=
    \oneover{\Z}\tr{
        e^{-(\beta-\tau) H}\
        \O_x\ e^{+H(t+\tau)}\ e^{-Ht}\adjoint{\O}_y
    }
    =
    \oneover{\sum_{a} e^{-E_a \beta}} \sum_{bc} e^{-E_b \beta} e^{-(E_c-E_b)\tau} z_{axb} z_{b\adjoint{y}a}
    \label{eq:spectral decomposition}
\end{align}
where we moved to the Heisenberg picture and inserted resolutions of the identity in the energy eigenbasis and defined the overlap factors
\begin{align}
    z_{axb} &= \left\langle a \middle| \O_x \middle| b \right\rangle
    &
    z_{b\adjoint{y}a} &= \left\langle b \middle| \adjoint{\O}_y \middle| a \right\rangle = z_{ayb}^*
    \label{eq:overlap factors}
\end{align}
where $a$ and $b$ label energy eigenstates.
In the low temperature limit $\beta\goesto\infty$ the sum is dominated by the lowest energy state; otherwise thermal artifacts may be seen.
In the low temperature and late-(euclidean-)time limit $\tau\goesto\infty$, the correlator's $\tau$ dependence gives the energy gap between the ground and the first excited state (with the observable's quantum numbers), as all the heavier states decay more quickly.

Hence, we can extract the single-particle/single-hole  spectrum by e.g. setting $\adjoint{\O}_x = \adjoint{a}_x$. $C_{xy}(\tau)$ is quadratic in the volume. However, after diagonalization one obtains only $O(V)$ single-particle eigenfunctions. For the small lattices we study here, it suffices to project $x$ and $y$ to the same irreducible representation of the lattice symmetry in order to diagonalize $C$.

We can also calculate correlation functions between composite operators at additional computational expense.
The composite operators we consider are the number operators $n^a_x$ and $n^b_x$, their sum the total number operator $n_x$, their difference the charge operator $\rho_x$, spin raising and lowering operators $S^\pm_x$, and the spin operators $S^i_x$ (where $i$ runs over all 3 spatial directions); we also consider the two doubly-charged local bilinears.
In \Appref{bilinears} we detail the operators, the correlation functions we measure and how to construct conserved quantities from them.



\section{Algorithm}\label{sec:algorithm}

We gave an extensive overview of our application of Hybrid Monte Carlo (HMC) to the Hubbard model in \Ref{Wynen:2018ryx}.
In all our reweighting-only examples we run HMC in a standard way, performing the Metropolis accept-reject step according to the real part of the action.
In the rest of this section we detail how we incorporate learnifolds into HMC.

In \Secref{flow} we provide a summary of why integrating over a manifold given by holomorphic flow is advantageous for reducing the sign problem.
Then in \Secref{networks} we show how we use machine learning to quickly compute the learnifold, over which we will integrate. Afterwards we explain how to incorporate the learnifold into HMC and, finally, comment on our update scheme's ergodicity in \Secref{hmc}.

\subsection{Holomorphic Flow}\label{sec:flow}
Holomorphic flow is a generalization of the steepest descent method to multi-dimensional complex space.
Given a holomorphic functional, in our case the action $S$, of the $N$-dimensional complex $\phi$, the flow equations are
\begin{equation}\label{eq:flow eq}
  \od{\phi}{\tau_f} = \pm\left(\pd{S[\phi]}{\phi}\right)^*\ ,
\end{equation}
where $\tau_f$ is the \emph{flow time}.
A minus sign indicates \emph{downward} flow, while a plus sign \emph{upward} flow.
Splitting the components $\phi_i=\phi^R_i+i \phi^I_i$ and the action $S[\phi]=S^R[\phi]+i S^I[\phi]$ into their real and imaginary parts, yields
\begin{align}
  \od{\phi^R}{\tau_f}&=\pm\pd{S^R}{\phi^R_i}=\pm\pd{S^I}{\phi^I_i}\\
  \od{\phi^I}{\tau_f}&=\pm\pd{S^R}{\phi^I_i}=\mp\pd{S^I}{\phi^R_i}\ ,
\end{align}
which are the Cauchy-Riemann equations.
The equations containing $S^R$ essentially implement the gradient flow, whereas the other equations are Hamilton's equations for the imaginary part of the action.
That is, $S^I$ remains a constant of motion during the flow.
It is easily seen that downward holomorphic flow is the generalization of gradient flow, or steepest descent, for real fields.

The critical points $\hat{\phi}_{cr}$ of $S[\phi]$ are vectors of complex numbers and satisfy
\begin{align}
  \pd{S[\phi]}{\phi}\bigg|_{\hat{\phi}_{cr}} = 0\ .
  \label{eq:critpflow}
\end{align}
The $\hat{\phi}_{cr}$ are saddle points; the associated \emph{Lefschetz thimble} is the manifold in $\Complexes^N$ which flows to the given critical point under downward flow, while the \emph{dual thimble} flows to the critical point under upward flow.
Models with many variables have many critical points and associated thimbles that must be integrated over to produce the same result as the integral on the real manifold.

Flowing the integration region of $\phi\in \Reals^N$ to the manifold(s) $\hat{\phi}$ eliminates the sign problem since $S^I$ is constant on each manifold.
Because there are no poles in the integration kernel, Cauchy's integral formula guarantees that the integral over the thimbles $\hat{\phi}\in\Complexes^N$ will be exactly equal to the original integral over $\phi\in\Reals^N$,
\begin{equation}
  \int \Dif \phi\, e^{-S[\phi]}
  =
  \sum_\sigma
    e^{-i S^I[\hat{\phi}_{cr,\sigma}]}
    \int\Dif \hat{\phi}_\sigma\, e^{-S^R[\hat{\phi}_\sigma]}\ ,
\end{equation}
where $\sigma$ runs over included thimbles and $\hat{\phi}_{cr,\sigma}$ is the critical point associated with this thimble.
The number of thimbles, their critical points, and their relative weights depend on the lattice action and the original integration region (typically $\Reals^N$).
Not all the thimbles in $\Complexes^N$ are included in the sum; only those that are required to preserve the integral's homology class.

Thimbles do not cross each other, but are connected at places where the action diverges so that the integration kernel vanishes.
The gaussian part of the action diverges when $|\hat{\phi}|\to\infty$; the fermionic part of the action diverges when $\det (M[\hat{\phi}]M[-\hat{\phi}])=0$.
We call these zero-weight places in the complex space \emph{neverland}, as they never appear in an importance-sampling scheme.

Because the locations of the needed critical points and their associated thimbles are not known \emph{a priori}, their determination requires extensive numerical resources; an analytic determination would amount to a solution of the lattice model.
Instead we follow Alexandru, Basar, and Bedaque~\cite{Alexandru:2015xva} and use the fact that the thimbles are fixed points of the flow to our advantage: we flow only a modest amount to get an integration manifold that approaches the set of relevant thimbles, not solving the sign problem entirely but alleviating it to the point where standard reweighting techniques are sufficient to address any remaining sign problem.

To ensure our method remains exact, it is important that our ultimate integration manifold is in the same homology class as the original.
Because the holomorphic flow preserves the homology class, the manifold resulting from any finite flow time is in the right class.
Moreover, because the flow's fixed points are thimbles, a finite-flow manifold approaches those thimbles that contribute to the integral~\cite{Alexandru:2015xva}; we need not analytically decide which thimbles to include---the flow discovers this automatically.

One point on the original manifold flows to a thimble's critical point, and only a vanishing neighborhood around that point flows to the rest of the thimble---most of the original manifold flows to \emph{neverland}---the place where thimbles meet and vanish in the integration kernel, typically due to zeros of the fermion determinant.
These zeros act as attractors, and most configurations flow to these zeros after a finite amount of flow time~\cite{Kanazawa:2014qma,Alexandru:2015xva,Mori:2017zyl}.
Because the thimbles must meet at these zeros, attempting an update method like HMC on the thimbles themselves in a naive way will be obstructed by these zeros, causing an ergodicity problem.
However, by restricting ourselves to modest flow times, we can ensure that the manifold will not touch these zeros.
We provide some visual evidence of ergodicity in \Secref{triangle ergodicity} and reproduce a wide range of exactly-known correlation functions.
So, deciding how much to flow is a balancing act---one hopes to flow enough that the sign problem is alleviated but not so much that an ergodicity problem emerges.

Any transformation of integration variables, including the one provided by the flow, comes with an associated Jacobian, and this Jacobian must be included in the Monte Carlo weight (or incorporated by reweighting) for a correct method.
There are flow equations for the Jacobian as well~\cite{Alexandru:2015sua},
\begin{equation}\label{eq:jacobian}
\od{J}{\tau_f}(t)=\left[H[\phi(t)] J(t)\right]^*\ ,
\end{equation}
where $H$ is the Hessian,
\begin{equation}\label{eq:hessian}
    H_{ij}[\phi]
    \equiv
    \left.\frac{\partial^{2} S}{\partial \phi_{i} \partial \phi_{j}}\right|_{\phi}
    =
        \left.\left(
                \tr{M\inverse (\partial_i M) M\inverse (\partial_j M)}
            -   \tr{M\inverse \partial_i\partial_j M}
        \right)\right|_{+\kappa,+\phi}
    +   \left(\kappa, \phi \goesto -\kappa, -\phi\right)
    +   \frac{\delta_{ij}}{\tilde U}
    \ .
\end{equation}
Flowing the Jacobian is the most time-consuming aspect of this calculation and thus the reason for considering neural networks, as we discuss in the following section.

To perform the numerical integration of the flow equations~\eqref{flow eq}, and when necessary of the Jacobian~\eqref{jacobian}, we use a 4$^{th}$ order adaptive Runge-Kutta method.
Because the flow preserves the imaginary part of the action, we monitor the latter during the numerical integration and adjust the integration stepsize to keep deviations within a prescribed tolerance.


\subsection{Networks}\label{sec:networks}

\subsubsection{Architecture}

We use feed-forward neural networks of dense layers to tackle the sign-problem.
Like Refs.~\cite{Alexandru:2017czx,Mori:2017nwj} these networks produce the imaginary part of a field configuration from the real part and keep the latter fixed.
The transformation into complex space is thus
\begin{align}\label{eq:network transform}
  \tilde{\phi} = \phi + i\ \NN(\phi),
\end{align}
where $\NN$ is the neural network.
In addition to the ability of the network to avoid dealing with complex numbers directly, \Ref{Alexandru:2017czx} points out two advantages this formulation enjoys over $\tilde{\phi} = \NN(\phi)$: it might ameliorate the ergodicity problem that flowing can induce (for an enlightening discussion and illustration see their Figure 2); and it tends to yield a more stable Jacobian (Figure 3 of  \Ref{Alexandru:2017czx}).

We encode $\phi$ as a spacetime vector and use neural networks with a single hidden layer, consisting of twice the number of neurons than the input and output layers.
The hidden layer has a Softplus activation function
\begin{align}
  \text{Softplus}(x) = \log(1 + \exp(x)),
\end{align}
and the output layer has none.
We found this to be sufficient for all cases we tested.
Wider or deeper networks as well as convolutional layers or different activation functions did not yield significantly better results.
It remains to be seen whether this setup scales to larger spatial lattices.

Neural networks of the chosen architecture can be evaluated very efficiently, which is, unfortunately, only half of the story.
We also need to compute the Jacobian determinant for the change of variables, which will appear in the transformed integral~\eqref{integral transform}.
Given~\eqref{network transform}, it is
\begin{align}\label{eq:network jacobian}
  \det J[\phi] = \det \bigg(\one + i\pd{\NN_i(\phi)}{\phi_j}\bigg)\ .
\end{align}
This operation is $\mathcal{O}\big({(N_sN_t)}^3\big)$ (cubic in the spacetime volume), because both the determinant and matrix multiplications in the derivative need to be performed with a general algorithm for dense matrices.
No speedup seems possible for vanilla dense networks as the weight matrices are unconstrained.
We still find our network-based transformation to significantly outperform flow-based transformations in terms of run time, however.
But this approach does not scale to larger lattices.
It is possible to improve on the scaling by estimating the Jacobian as in Ref.~\cite{Alexandru:2017czx} which uses a different network that produces only one component of $\tilde{\phi}_I$ at a time.
Alternatively, one can use coupling layers which were designed to have simple Jacobians.
This requires complex valued networks in our case, however, as described in Appendix~\ref{sec:coupling layers}.

We implemented the neural networks and training procedures using PyTorch~\cite{NEURIPS2019_9015} and used Isle~\cite{isle03} to implement Monte-Carlo for the Hubbard model.
\Secref{hmc} describes the HMC scheme we used to accommodate neural networks.

\subsubsection{Training procedure}
We train our models using a supervised approach.
To that end, we generate random real configurations $\phi$ and flow them upward according to Eq.~\eqref{flow eq} for a fixed flow time to $\hat{\phi}$.
It is non-trivial how to generate useful data and we provide explicit details on our approach in \Secref{training data}.
Once the data is generated, we use $\Re\, \hat{\phi}$ as inputs and $\Im\, \hat{\phi}$ as target outputs to train the networks.
We do so by using the Adam algorithm~\cite{Kingma:2014ada} to minimise a smooth L1 loss
\begin{align}
  \text{loss}(x, y) = \frac1{n}\sum_i^n
  \begin{cases}
    {(x_i - y_i)}^2 / 2, & \text{for} \abs{x_i-y_i} < 1\\
    \abs{x_i - y_i} - 1/2, & \text{otherwise}
  \end{cases}
\end{align}
as defined in PyTorch~\cite{NEURIPS2019_9015}.

We know our action has exact symmetries (see \Ref{Wynen:2018ryx} for a listing), comprising transforms $T$ that change field configurations but leave the action invariant, $S[T\phi] = S[\phi]$.
A uniformly-flowed manifold exhibits many of those symmetries, the most obvious being the temporal and spatial translation symmetries. Naturally it is desirable that our neural networks preserve as many of these symmetries as possible.

\Ref{Alexandru:2017czx} accomplishes this by training a network that takes a whole spacetime vector and produces only the flowed configuration at the spacetime origin and using translational invariance to construct the other vector elements.
This technique suffers a number of constraints, most notably, it requires full translational invariance of the lattice.
This is not the case for many interesting non-bipartite lattices, however, as they are not necessarily vertex-transitive.
In such a case a network with a single output is not enough because spatial symmetries cannot take every element to every other.\footnote{It may still be possible for the network to produce a much smaller vector --- for example, the size of a unit cell, or just a single timeslice.}

As described above, our neural networks produce a full spacetime vector as output and we encode symmetries by augmenting the training data.
One symmetry that all lattices have is temporal translational invariance.
We thus train our networks not only on the configurations generated according to \Secref{training data} but also on all possible temporal shifts of the input and target output configurations.
This effectively increases the size of the data set by a factor of $N_t$.
The lattices considered in this work, triangle and tetrahedron, are vertex-transitive and we tried augmenting the training data by all possible spatial permutations but did not find an improvement.
Adding temporal translations proves very useful for increasing model quality, however.

Every model was trained on minibatches of size 16 drawn from 1000 random configurations plus temporal translations.
The only exception is the tetrahedron with $N_t = 32$ whose model was trained with 4000 configurations.
We found this larger number necessary to train the network well.
For larger $N_t$, 1000 configurations were enough for the tetrahedron which might be due to the higher number of added configurations from augmentation by temporal translations.

\subsubsection{Training: Data}\label{sec:training data}

Several different methods have been used to generate training data.
\Ref{Mori:2017nwj} uses HMC with a partially trained model to generate data for the next iteration of training.
This is expensive as new configurations have to be generated every time model hyperparameters are changed.
Because it is self-reinforcing, it is possible that such an approach can find and remain in a local minimum or overfit a part of the phase space.

\Ref{Alexandru:2017czx} uses HMC without a neural network to generate training data before fitting a model.
In order to achieve large phase space coverage, several MC ensembles with different temperatures were generated.
This approach needs to generate data only once per parameter set and therefore allows for faster tuning of hyperparameters.
It is still susceptible to autocorrelations, though, meaning that a large number of configurations might be needed for sufficient phase space coverage.

One of the goals of this work is to simplify the generation of training data in order to develop a more scalable approach.

\begin{figure}[ht]
  \centering
  \includegraphics[width=0.8\textwidth]{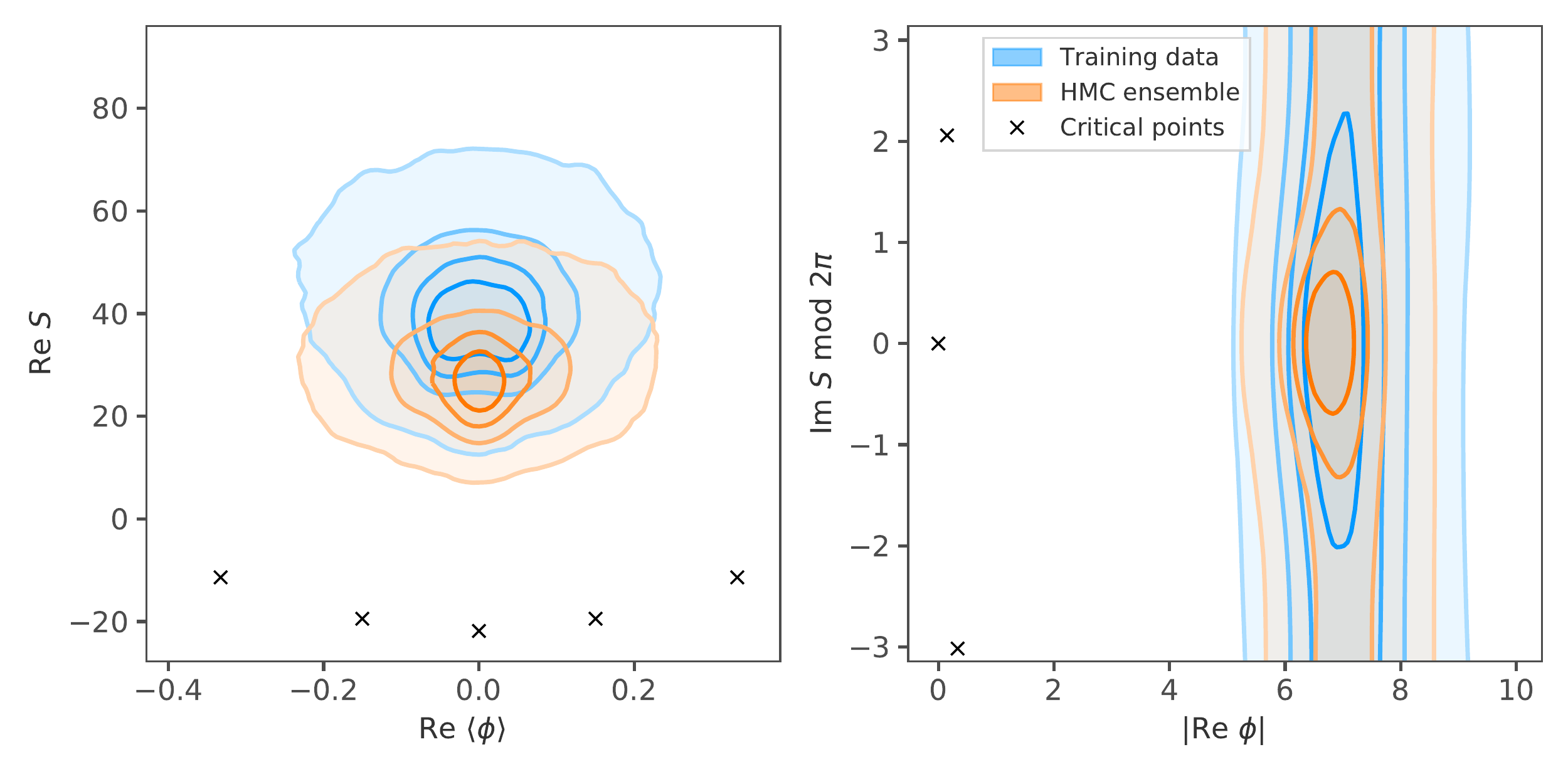}
  \caption{
    Distributions of configurations used for training data (blue-shaded region) and an ensemble produced via HMC with a neural network (orange-shaded region) trained on this data.
    The system is a triangle with $N_t=32$, $\kappa\beta=6$, and $U/\kappa=3$.
    The distributions were estimated from $10^5$ training and $10^6$ HMC configurations.
    The black crosses mark critical points of spacetime-constant $\phi$.\label{fig:training data}}
\end{figure}

One thimble of the Hubbard model on non-bipartite lattices is easy to find --- the image of $\Re\ \phi \equiv 0$ under the flow.
It connects to a critical point $\hat{\phi}_{\text{cr}} \equiv ic$ for some $U$, $\beta$, and $N_t$ dependent value $c < 0$.
In concordance with literature, we call this the ``main'' thimble.
There is an ongoing discussion whether a single thimble \Ref{Cristoforetti:2012su} or multiple thimbles are require to solve the integral~\cite{Alexandru:2015xva,Alexandru:2015sua,Kanazawa:2014qma,Tanizaki:2016lta}.
To be on the safe side, we, therefore, identified additional thimbles, all with spacetime-constant $\hat{\phi}_\text{cr}=z\in\Complexes$, but close to the main thimble.
\Figref{training data} shows the critical points for a sample system.

We produced training data by first generating random field configurations close to the critical points that we found and then flowing them for a short, fixed flow time.
In order to reduce the required flow time, we sampled the initial configurations on the plane tangent to the main thimble in the critical point ($\phi = \phi^R + ic$); this is referred to as the ``tangent plane'' from now on.
Models trained on these configurations for a triangle lattice can sometimes reduce the sign problem but are prone to producing learnifolds with a wrong homology class.
This problem becomes worse the larger $N_t$ is and we did not obtain any successful models for the tetrahedron.
The reason is the following.

Empirically, we found that the dependence of the typical set of the probability distribution $W$ on the vector norm $\abs{\Re\, \phi}$ is almost the same as that of a normal distribution.
In high dimensions, the typical set of a gaussian, that is the region of phase space that contains most of the probability, is a thin hyperspherical shell around the origin.
Importance sampling, by definition, generates fields that lie in the typical set.
We can thus use HMC to visualize the set, an example is shown in orange in \Figref{training data}.
The critical points of all thimbles that we used are shown as crosses in the figure.
It can be seen that those points are close to the origin and far away from the typical set.
Therefore, it should not be surprising that neural networks trained on fields near the critical points do not generalize well to the typical set and yield poor performance of HMC.\@

The approach used in this work is motivated by the dominance of the gaussian part of the action.
We draw $\Re\ \phi$ randomly from a gaussian and set $\Im\ \phi$ to be on the tangent plane of the main critical point.
These $\phi$ are then flowed upwards.
This method has two numerical parameters that need to be tuned.
One is the width of the gaussian $\sigma$, the other is the flow time.
Plots like \Figref{training data} provide good estimates for the quality of training data.
The better the overlap between training data and HMC in these projections the better the neural network performs.
Fortunately, the distributions explored by HMC with and without neural networks are largely identical in these projections for well trained models.
It is thus sufficient to sample configurations on the real plane in order to estimate the HMC distribution --- even though the figure shows an ensemble generated \textit{with} a network.
The overlap shown in the figure, while not perfect, is enough and the figure shows data from a successful run with reduced sign problem shown in more detail below.

Samples drawn from a normal distribution of fixed width $\sigma = \sqrt{\Utilde}$ have poor overlap in figures like \Figref{training data}.
To remedy this, we draw the width of every sample randomly from a uniform distribution $\mathcal{U}$
\begin{align}
  \sigma \sim \mathcal{U}\Big(\sqrt{\Utilde} / x,\; \sqrt{\Utilde}\,\Big)
\end{align}
and then draw $\phi$ from a normal distribution $\mathcal{N}$ with that width: $\Re\ \phi \sim \mathcal{N}(0, \sigma)$.
The factor $x$ is chosen such that the overlap of HMC ensemble and training data is maximized.

Similarly, the other parameter, the flow time $\tau_f$, needs to be chosen such that plots like \Figref{training data} show good overlap.
Additionally, one needs to strike a balance between reducing the sign problem and avoiding ergodicity problems as described in \Secref{flow}.
It is easy to show that for spacetime-constant $\phi$ the gradient of the action is invariant under simultaneous $N_t \to \alpha N_t$ and $\phi \to \phi / \alpha$.
The starting position for flow, which is on the main tangent plane, has, therefore, the same $N_t$ dependence.
We use this scaling for the flow time as well and choose it to be
\begin{align}
  \tau_f^\text{max} = 0.1 \times 16 / N_t
\end{align}
in all cases.
As with $\sigma$, this scaling was found purely based on numerical experiments.

We found significant improvements in both the time required to make training data and the performance of the final neural networks by not requiring a fixed flow time.
When using holomorphic flow directly within MC, it is necessary to fix the flow time in order for all configurations to be on the same manifold.
In our case, the neural network ensures this regardless of how the training data was created.
Many configurations generated from a gaussian quickly flow into neverland and flowing becomes numerically unstable.
It is thus impossible to reach the targeted flow time.
Instead of discarding these configurations, we monitor the integrator for stability, abort early in such a case, and add the last configuration that could be reached in a stable way to the set of training data.
However, it does not make sense to use configurations that could only be flowed for very short times as those provide little information to the neural network.
We thus require a minimum flow time of
\begin{align}
  \tau_f^\text{min} = 0.04 \times 16 / N_t
\end{align}
for all training configurations.


\subsection{HMC}\label{sec:hmc}

Rather than implementing HMC on the curved manifold of Lefschetz thimbles defined through holomorphic flow~\cite{Fukuma:2019uot} or learnifolds, we pull back to the real plane before performing molecular dynamics, as will now be explained.

Given a transformation into complex space as implemented by holomorphic flow or neural networks, we need to incorporate that transformation into Hybrid Monte-Carlo.
Let
\begin{align}
  f: \Reals^N \to \Complexes^N, \phi \mapsto \tilde{\phi} \qquad\text{and}\quad J_{ij}[\phi] \equiv \pd{f_i(\phi)}{\phi_j}\ ,
\end{align}
with $\mathcal{M} \equiv f(\Reals^N)$ the image of $f$.
We want to calculate observables by integrating over the manifold $\mathcal{M}$ because with suitably chosen $f$ the sign problem is alleviated on that manifold.
Since $\mathcal{M}$ is in the same homology class as $\Reals^N$ and the integrand is regular in $\phi$, by Cauchy's theorem the expectation value of an observable $\O$ is
\begin{align}
  \langle\O\rangle = \frac1{\mathcal{Z}_{\Reals^N}} \int_{\Reals^N}\Dif \phi\, \O[\phi] e^{-S[\phi]} = \frac1{\mathcal{Z}_{\mathcal{M}}} \int_\mathcal{M}\Dif \tilde{\phi}\, \O[\tilde{\phi}] e^{-S[\tilde{\phi}]}\ .
\end{align}
We do not know $\mathcal{M}$, however, and can thus not evaluate the integral on the right hand side directly.
Instead, we parameterize it using $f$.
That is, we perform a transformation of integration variables:
\begin{align}\label{eq:integral transform}
  \frac1{\mathcal{Z}_{\mathcal{M}}} \int_\mathcal{M}\Dif \tilde{\phi}\, \O[\tilde{\phi}] e^{-S[\tilde{\phi}]}
  &= \frac1{\mathcal{Z}_{\Reals^N}} \int_{\Reals^N}\Dif \phi\, \O[\tilde{\phi}(\phi)] e^{-S[\tilde{\phi}(\phi)]} \det J[\phi]\ .
\end{align}
Now define the effective action as
\begin{align}
  \Seff[\tilde{\phi}(\phi)] \equiv S[\tilde{\phi}(\phi)] - \log \det J[\phi]\ .
\end{align}
We can estimate the integral stochastically by generating an ensemble $\{\phi \sim \exp(-\Re \Seff[\tilde{\phi}(\phi)])\}$, where we use the real part of $\Seff$ as per the reweighting procedure described in \secref{reweighting}.
This effectively produces an ensemble $\{\tilde{\phi} = f(\phi) \sim \exp(-\Re S[\tilde{\phi}])\}$ on which the observables $\O$ can be measured.

We use HMC to generate the ensembles.
To this end, we augment the integral by multiplying with a one in the form of an integral over the artificial conjugate momentum $p$ such that
\begin{align}
  \langle\O\rangle = \frac1{\mathcal{Z}} \int\Dif \phi \Dif p\, \O[\tilde{\phi}(\phi)] e^{-i\Im \Seff[\tilde{\phi}(\phi)]} e^{-H[p, \tilde{\phi}(\phi)]}, \quad H[p, \tilde{\phi}(\phi)] \equiv \frac{p^2}{2} + \Re \Seff[\tilde{\phi}(\phi)]\ .
\end{align}
All integrals are to be understood as integrating over $\Reals^N$ from now on.
We can encapsulate all dependencies on $f$ in a modified HMC algorithm such that it produces configurations on $\mathcal{M}$ suitable for measurements.
The following summarizes the algorithm:\\
\begin{center}
  \begin{tabular}{ll}
    \multicolumn{2}{l}{\underline{\textsf{complexified HMC}}\qquad {\color{gray}in: }$\tilde{\phi}$\quad{\color{gray}out: }$\tilde{\phi}'$}\\[6pt]
    \qquad $\phi \leftarrow f^{-1}(\tilde{\phi})$ & \quad {\color{gray} \hash transform to $\Reals^N$}\\
    \qquad $p \leftarrow \mathcal{N}_{0,1}$ & \quad {\color{gray} \hash draw random momentum}\\
    \qquad $\psi',\, p' \leftarrow \textsf{molecular\_dynamics}(\phi,\, p)$ & \quad {\color{gray} \hash generate candidate on $\Reals^N$}\\
    \qquad $\tilde{\psi}' \leftarrow f(\psi')$ & \quad {\color{gray} \hash transform to $\mathcal{M}$}\\
    \qquad $\tilde{\phi}' \leftarrow \textsf{accept\_reject}(H[p',\,\tilde{\psi}'], H[p,\,\tilde{\phi}])$ & \quad {\color{gray} \hash pick new field}
  \end{tabular}
\end{center}
A new configuration is obtained from an old one by first transforming the old complex field $\tilde{\phi}$ to the real plane\footnote{The inverse transformation $f^{-1}$ can be unstable and/or expensive to evaluate. But there is no need to perform this calculation if we just keep track of $\phi$ as well as $\tilde{\phi}$.} where it is then updated using molecular dynamics (MD), or any other suitable updating scheme, such as the large jumps explained in \Ref{Wynen:2018ryx}.
The new candidate field $\psi$ is then transformed to $\mathcal{M}$ where it is accepted or rejected using a Metropolis-Hastings step.

It remains to prove that this algorithm produces a Markov Chain.
For this, we adopt the picture that we are producing real fields $\{\phi \sim \exp(-\Re \Seff[\tilde{\phi}(\phi)])\}$.
Such a proof is equivalent to showing that $\{\tilde{\phi}\}$ is a Markov Chain sampled from $\exp(-\Re S[\tilde{\phi}])$.
We do so by first proving detailed balance
\begin{align}\label{eq:detailed balance}
  \mathds{P}(\phi) \mathds{P}(\phi'|\phi) = \mathds{P}(\phi') \mathds{P}(\phi|\phi')
\end{align}
with (ignoring normalization factors)
\begin{align}
  \mathds{P}(\phi) &= e^{-\Re \Seff[\tilde{\phi}(\phi)]}\\
  \mathds{P}(\phi'|\phi) &= \int \Dif p' \Dif p\, e^{-p^2/2} \mathds{P}_{\text{MD}}(p',\phi'|p,\phi) \mathds{P}_{\text{a/r}}(p',\tilde{\phi}(\phi')|p,\tilde{\phi}(\phi))\ .
\end{align}
$\mathds{P}_{\text{MD}}$ and $e^{-p^2/2}$ are the same molecular dynamics and gaussian probabilities as in standard HMC.\@
The prior $\mathds{P}(\phi)$ and accept/reject
\begin{align}
  \mathds{P}_{\text{a/r}}(p',\tilde{\phi}(\phi')|p,\tilde{\phi}(\phi)) = \min\left(1,\; \frac{\exp\left(-\frac{p'^2}{2}-\text{Re}\Seff[\tilde{\phi}(\phi')]\right)}{\exp\left(-\frac{p^2}{2}-\text{Re}\Seff[\tilde{\phi}(\phi)]\right)}\right)
\end{align}
probabilities use the effective action which encapsulates $\tilde{\phi}$.
Thus the proof of detailed balance~\eqref{detailed balance} proceeds as usual for HMC.\@
The only ingredient missing to fully prove correctness of our algorithm is a proof of ergodicity.
Such a proof is generally not available even for standard HMC.\@
We thus rely on \emph{a posteriori} analyses to verify ergodicity.
Certainly with a long flow time we expect many configurations to flow to neverland, creating large zero-probability regions that separate important islands of configurations.
However, if we only flow a little, few if any configurations flow to neverland and the manifold of integration is not partitioned.
Our ability to reproduce exact results in \Secref{results} suggests that our method successfully explores fields whose images are near different thimbles.
A well trained network inherits these properties from the flowed manifold of its training data.



\section{Results}\label{sec:results}

To demonstrate the efficacy of the neural network methods we will explore small lattices, leaving larger lattices for future work.  In particular, we here consider the triangle and tetrahedron, both maximally connected and therefore, where we expect the worst sign problem.

\subsection{The Triangle}\label{sec:triangle}

Three spatial sites is the smallest nontrivial non-bipartite graph we might study---the two-site problem is bipartite and the one-site problem has no hopping at all.
We studied those problems extensively using the lattice methods applied here in \Ref{Wynen:2018ryx}.

\begin{figure}[t!]
	\centering
		\includegraphics[width=\columnwidth]{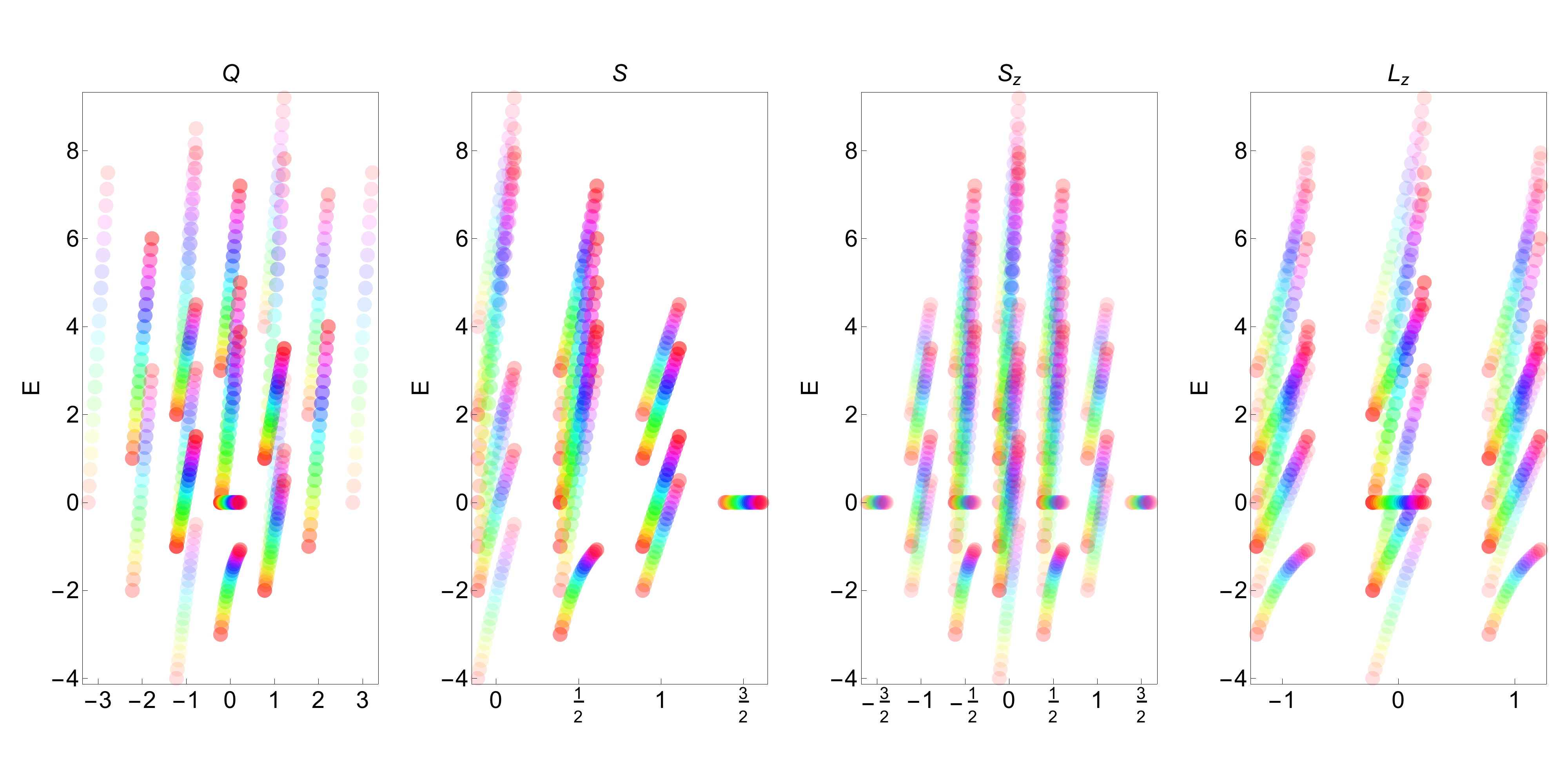}
	\caption{The entire spectrum of the Hubbard model on a triangle as a function of $U/\kappa$, changing from 0 to 5 in 20 equal steps, with colors corresponding to different $U/\kappa$.
    Energy eigenfunctions carry $Q$, $S$, $S_z$ and $L_z$ quantum numbers; each panel projects the five-dimensional space onto one quantum number and the energy.
    Each $U/\kappa$ is slightly offset, so that each point in the spectrum for $U/\kappa=0$ is to the left of that point's quantum numbers, while $U/\kappa=5$ is to the right.
    One translucent point is plotted per state; more opaque circles correspond to more highly degenerate states.
    The ground-state manifold qualitatively changes at $U^\triangle/\kappa=3.61775\ldots$, shown in {\color[hsb]{0.72355,1,1} a deep blue}.
    }
	\label{fig:triangle-spectrum}
\end{figure}

The hopping matrix is given by
\begin{equation}\label{eq:triangle K}
	h = \kappa \left( \begin{array}{ccc}
		0	&	1	&	1	\\
		1	&	0	&	1	\\
		1	&	1	&	0
	\end{array}\right),
\end{equation}
and the Hamiltonian has $D_3$ dihedral symmetry.
When diagonalized, the hopping matrix reveals one spatially uniform (trivial) irrep with eigenvalue $2\kappa$ and a dimension-two irrep with eigenvalue $-\kappa$.

The unitary site permutation operator $P$
\begin{align}
    P\; a_{x}\; \adjoint{P} &= a_{x+1}    &
    P\; b_{x}\; \adjoint{P} &= b_{x+1}    \nonumber\\
    P\; \adjoint{a}_x\; \adjoint{P} &= \adjoint{a}_{x+1}    &
    P\; \adjoint{b}_x\; \adjoint{P} &= \adjoint{b}_{x+1}    \label{eq:triangle permutation}
\end{align}
rotates the sites into one another (the indices on the ladder operators are understood mod 3). $P$ trivially commutes with the potential and nontrivially with the hopping \eqref{HO}.

The three irreducible single-particle destruction operators are
\begin{align}\label{eq:triangle single particle operators}
    \O_k    &=  \frac{1}{\sqrt{3}}\sum_{j=0}^2 e^{2\pi i j k / 3} a_j
\end{align}
which are labeled by $k=\pm1, 0$, which corresponds to their transformation properties,
\begin{equation}
    P \O_k \adjoint{P} = e^{2\pi i k / 3} \O_k,
\end{equation}
a spherical-tensor-like relation, and the hopping Hamiltonian can be decomposed
\begin{equation}
    H_0 = \kappa \left[\left(
            \adjoint{\O_{-1}}\O_{-1}
        +   \adjoint{\O_{+1}}\O_{+1}
        - 2 \adjoint{\O_{0}}\O_{0}
        \right) - (a \rightarrow b)
        \right],
\end{equation}
corresponding to the irreps described above.
We label the spatial permutation quantum number by $L_z=k$, since it corresponds to an angular momentum around the center of the triangle.

The spectrum for this system, which can be obtained from direct diagonalization of the Hamiltonian, consists of $4^3 = 64$ states in the entire Fock space.
When $U=0$, a single state with $Q=-1$, $S=0$, $S_z=0$ $L_z=0$ has the lowest energy, and a partner with $Q=+1$, $S=0$, $S_z=0$, $L_z=0$ has the highest energy.  Explicitly, the spectrum is not symmetric in $Q$, though of course the spectrum is symmetric in $S_z$ and $L_z$.
When $U$ becomes large, the lowest energy is shared by a degenerate quadruplet of states with $Q=0$, $S=1/2$, $S_z=\pm\half$ and $L_z=\pm1$.
At approximately $U^\triangle/\kappa = 3.61775$ the low-$U/\kappa$ ground state and high-$U/\kappa$ four-plet are degenerate.
When $U/\kappa$ is very large the spectrum nearly exhibits symmetry in $Q$.

Spectra, like those shown in \Figref{triangle-spectrum} and operator overlap factors \eqref{overlap factors}, can be used to directly calculate the single-particle correlator via the spectral decomposition \eqref{spectral decomposition}.
Using the single-particle operators \eqref{triangle single particle operators} we can also use Monte Carlo to compute the single-particle correlators numerically.


\subsubsection{Ergodicity and the Sign Problem}\label{sec:triangle ergodicity}

In the language of \Refs{Beyl:2017kwp,Wynen:2018ryx} we use the exponential $\alpha=1$ discretization.
As detailed in \Ref{Wynen:2018ryx} there is a formal ergodicity problem on bipartite lattices.
When the lattice is not bipartite the codimension-1 manifolds of exceptional configurations are reduced in dimension and there is no formal ergodicity problem.
First, we give a small toy problem confirming this claim and then reproduce the exact results obtained through direct diagonalization.
Then, we examine the statistical power for HMC alone as a function of $U/\kappa$ and $\kappa\beta$ and find that, for a given temperature, the sign problem is worst when $U/\kappa=U^\triangle/\kappa\approx3.61775$, the value where the vacuum changes character, as shown in \Figref{triangle-spectrum}.

\begin{figure}[htbp]
	\centering
	\includegraphics[height=3in]{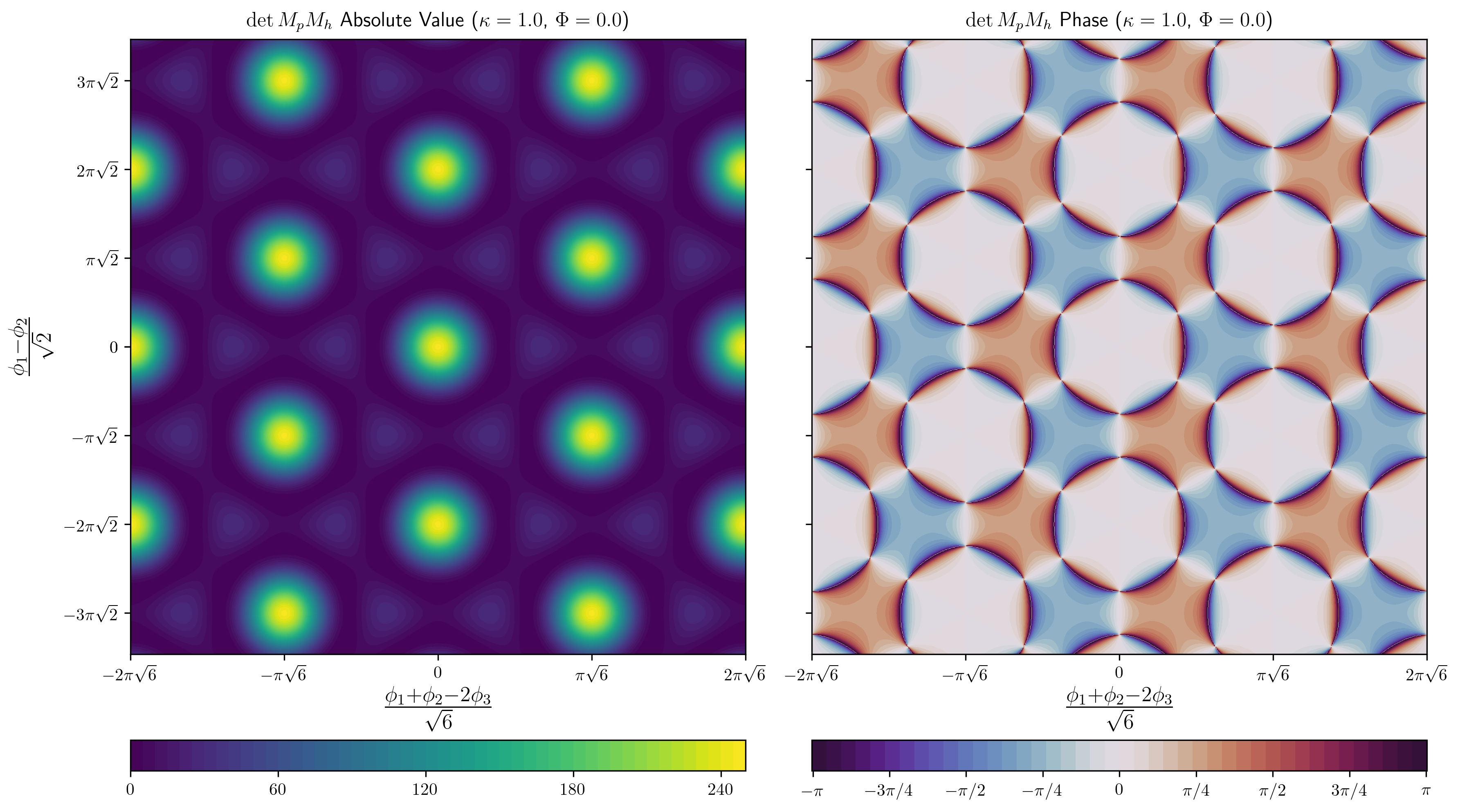}
	\caption{The absolute value and phase of \eqref{triangle-fermions} for $\kappa=1$ and $\Phi=0$, shown as a function of two directions in field space orthogonal to the $\Phi$ direction.  In the left panel, yellow indicates large absolute values and purple small ones.  In the right panel the color scheme is periodic; the exact zeros occur at the six points on each dark circle where the phase wraps around the point.}
	\label{fig:triangle-ergodicity}
\end{figure}

To visually appreciate that the codimension-1 manifolds that prevent HMC alone from being formally ergodic in the bipartite case are reduced in dimension, consider a problem with $\Nt=1$ and let $\phi_x$ live on spatial site $x$.
Then, the product of the fermion determinants is
\begin{align}
	\det M_p M_h = \frac{4}{9}\left[
			3 	\cos\left(\frac{1}{2}\Phi\right)
		+	\sum_{j=1}^3 \left(
				2	\cosh\left(\kappa + \frac{i}{2}(\Phi - 2\phi_j)\right)
				+	\cosh\left(2\kappa - \frac{i}{2}(\Phi - 2\phi_j)\right)
				\right)
	\right]^2
	\label{eq:triangle-fermions}
\end{align}
where $\Phi = \sum_x \phi_x$.
\Figref{triangle-ergodicity} shows the absolute value (left panel) and complex argument (right panel) of this determinant at $\Phi=0$ and two orthogonal combinations of the field variables.
The only zeros are where the lines of different phase meet---in the two-dimensional projection of the phase in \Figref{triangle-ergodicity}, points around which the phase winds.
As $\Phi$ changes, those points move but they never become extended.
So, even with an exponential  discretization, the codimension-1 zeros are reduced when the lattice is not bipartite; the fermion determinant allows free exploration the complex plane, rather than constraining it by the reality condition that arises in the bipartite case, as explained in \Ref{Wynen:2018ryx}.

\begin{figure}[htbp]
	\centering
		\includegraphics[width=.8\textwidth]{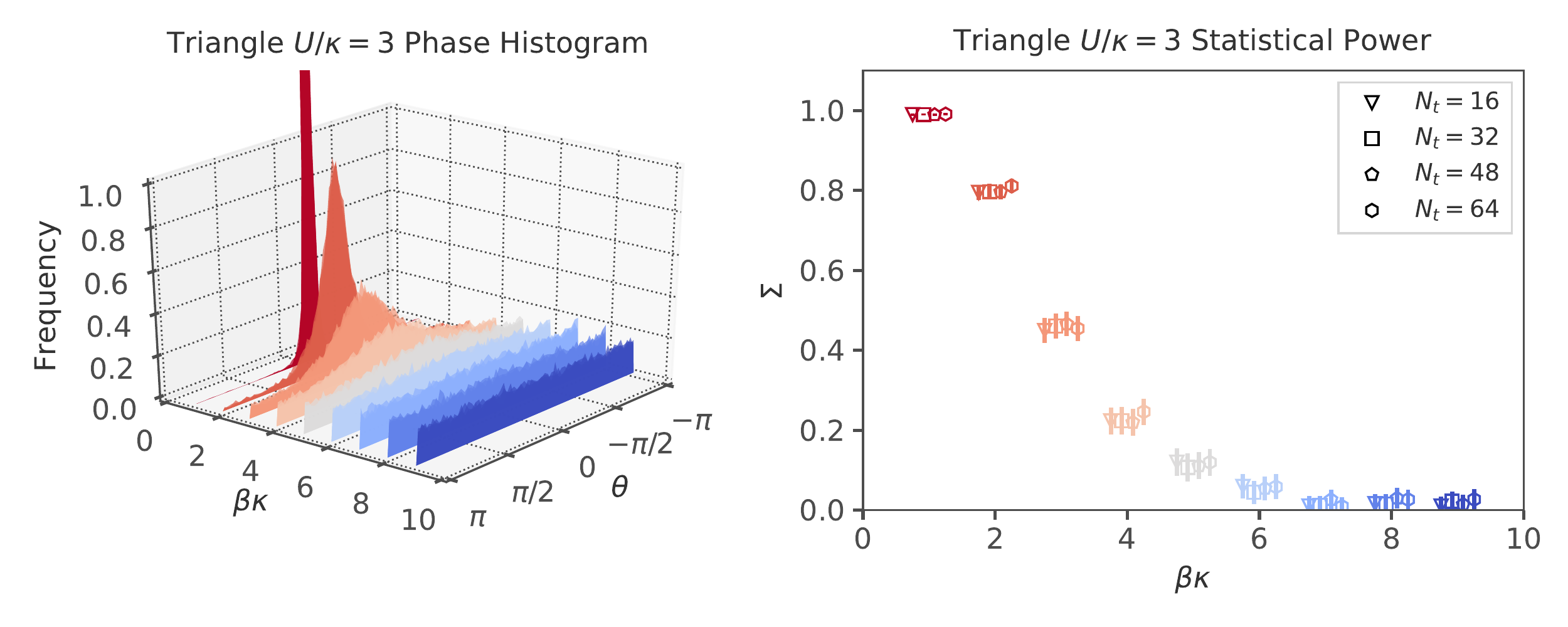}
		\includegraphics[width=.8\textwidth]{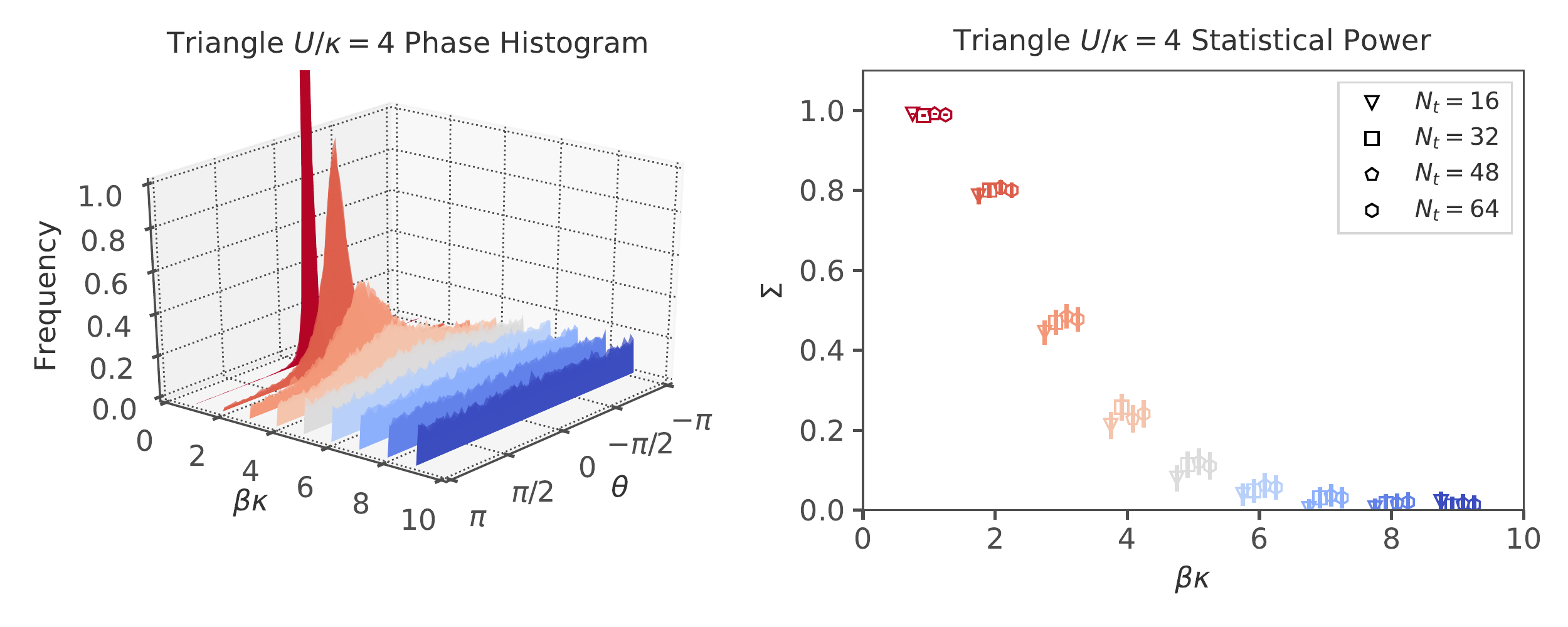}
	\caption{
        The histograms (left) of the phase for two different values of $U/\kappa$ as a function of $\beta$ and the corresponding statistical power (right) with bootstrap errors.
        In the right panel different $\Nt$ are shown from left to right: 16, 32, 48, and 64.
        }
	\label{fig:triangle phase histogram}
\end{figure}

The gaussian part of the weight encourages the fields to stay in the central mode.
When the gaussian becomes wider, more than one mode in \Figref{triangle-ergodicity} might become important.
There is no formal ergodicity problem, even in the exponential case, and HMC can take us from mode to mode.
However, at the mode boundaries, the phase changes rapidly, causing a sign problem.
In this small example, if the gaussian is wide enough, HMC can sample trajectories that near this rapid change we would expect to encounter a sign problem, but if the width were very narrow we would not.
In examples with more lattice sites and timeslices, the huge growth of phase space of configurations further from $\phi=0$ can counterbalance a narrow gaussian (the width is controlled by $U\beta/\Nt$), so it is \emph{a priori} unclear whether increasing $\Nt$ will help or hurt; since the action is extensive, one expects a sign problem exponentially bad with $\beta$.
However, increasing $\Nt$ also increases the number of variables and the odds that some are near the mode boundaries.

In \Figref{triangle phase histogram} we show the phase histogram (left panels) and the statistical power (right panels) of ensembles generated with HMC with real valued fields for two values of $U/\kappa$ as functions of $\beta\kappa$.
We generated 100,000 trajectories with one molecular dynamics time unit and the number of steps in the leapfrog integrator to give better than 75\% acceptance.
The histograms are actually results for multiple $\Nt$s superimposed, to show the very mild sensitivity to the discretization scale.
We also generated additional ensembles at $U/\kappa$ from one to nine in integer steps and $U^\triangle/\kappa$, and found that the sign problem modestly improved, for fixed $\kappa\beta$, for couplings further from $U^\triangle/\kappa$.
By analogy, we expect worse sign problems for critical points, where the system must tunnel between qualitatively different ground states.


\subsubsection{Results}\label{sec:triangle results}

Since we can exactly calculate the spectrum and overlap factors for this small problem, we can generate the exact, continuum-limit correlation functions according to the spectral decomposition \eqref{spectral decomposition}.  This provides us a means to directly check the accuracy of our NN method.

\begin{figure}
    \includegraphics[width=.45\textwidth]{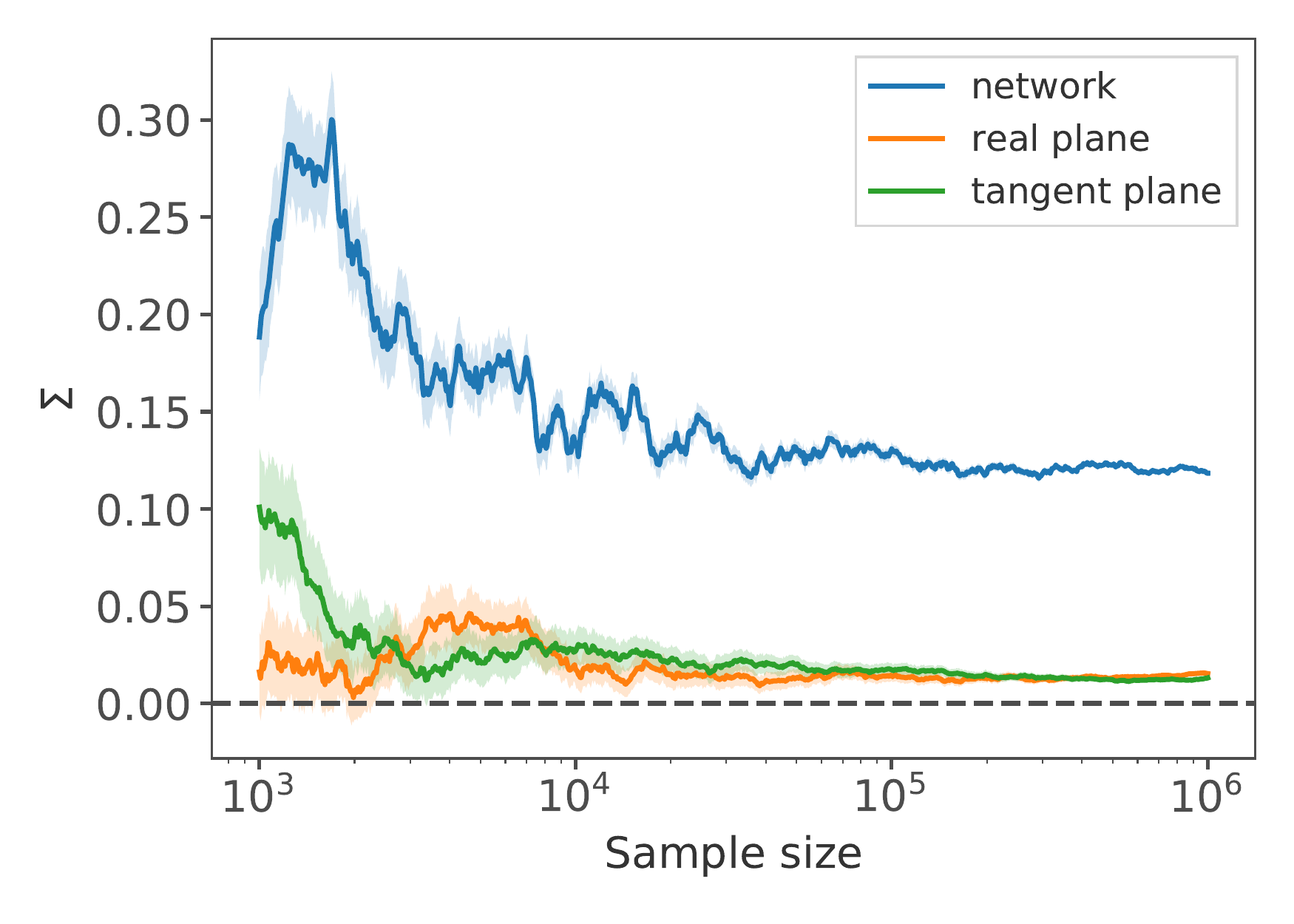}
    \includegraphics[width=.45\textwidth]{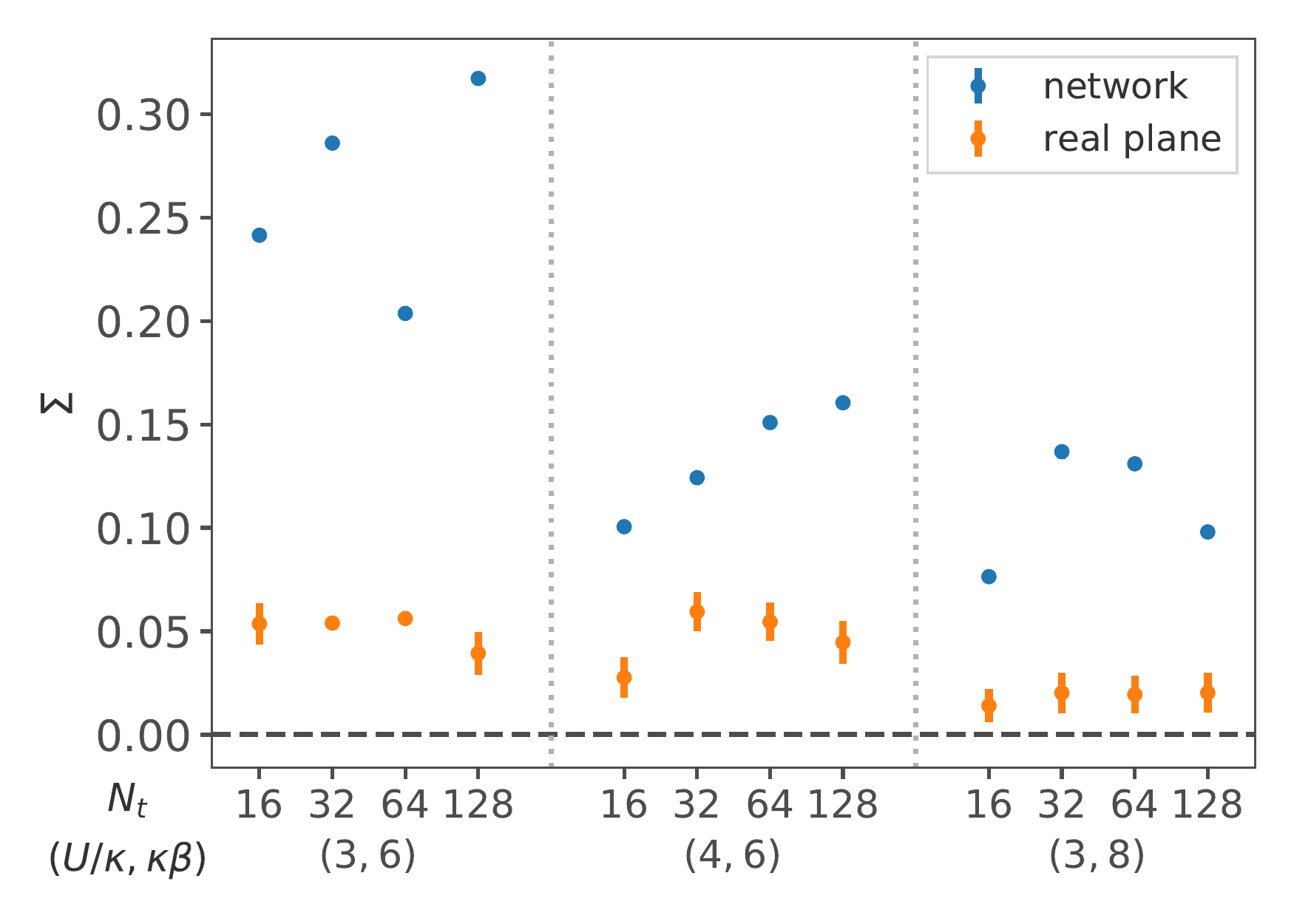}
    \caption{
    	Left panel shows statistical power as a function of sample size on a triangle lattice with $N_t = 64$, $U/\kappa = 3$, $\kappa\beta = 8$.
        Right panel shows statistical power for different parameters on a triangle lattice.
        Each point was estimated using $10^5$ configurations.\label{fig:triangle statpower overview}
        }
\end{figure}

We ran HMC in three different ways --- on the real plane, on the tangent plane of the main thimble, and on the learnifold.
The left panel of \Figref{triangle statpower overview} shows running averages of the statistical power as a function of Monte Carlo time for $U/\kappa = 3$ and $\kappa\beta = 8$.
After many configurations, $\Sigma$ on real and tangent planes converges to the same small but non-vanishing value, while the neural network produces a markedly greater statistical power.
Remember, an improved statistical power indicates an exponential reduction of the sign problem.
So, while the sign problem is not \emph{solved}, per se, we provide evidence here that it is significantly alleviated.

This chosen set of parameters shows the worst statistical power on real and tangent plane out of the sets we tested.
See the right panel of \Figref{triangle statpower overview} for a summary for different parameters.
The large variations in $\Sigma$ for neural network based calculations stem from different qualities of the trained models.
It should be possible to tune the networks better and thus increase statistical power.
These networks perform well enough, however.
Generally, we found that the sign problem is not a complete impediment on the triangle and a plain calculation on the real plane with reweighting can suffice.

\begin{figure}[ht]
  \centering
  \includegraphics[width=0.99\columnwidth]{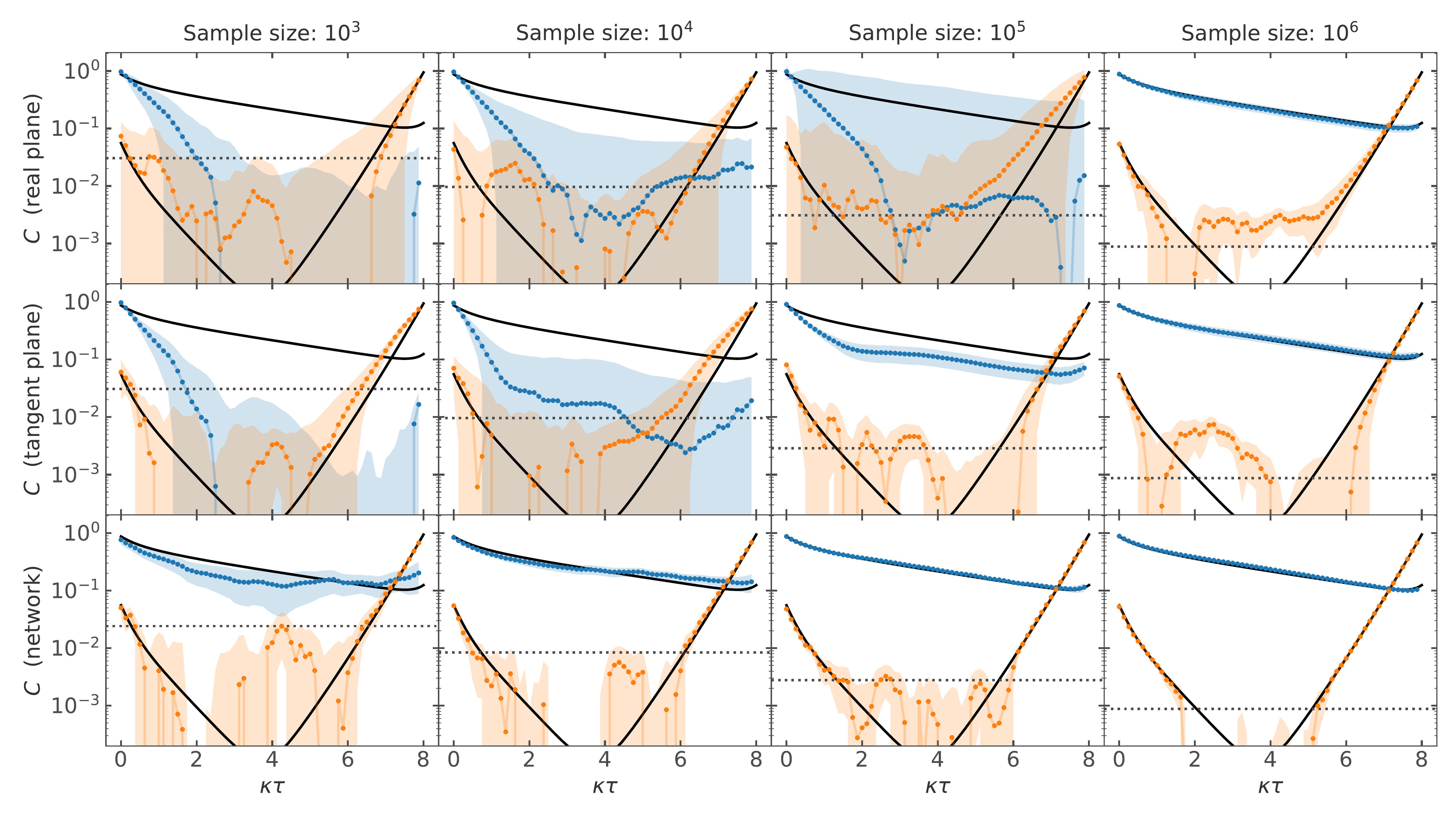}
  \caption{
    Single particle correlators $\langle a a^\dagger(\tau) \rangle$ on a triangle lattice with $N_t = 64$, $U/\kappa = 3$, $\kappa\beta = 8$.
    Each column shows correlators obtained from ensembles of the given number of configurations, while each row shows a different implementation of HMC.
    Sample sizes show the total number of Monte Carlo configurations but correlators were measured only on every $10^\text{th}$ configuration.
    The dotted lines are placed at $\max(C) / \sqrt{\text{Sample size}}$ and indicate the scale at which even a sign-problem-free method would show sizeable statistical fluctuations.\label{fig:triangle singleparticle convergence}
    }
\end{figure}

\begin{figure}[ht]
  \centering
  \includegraphics[width=.3\columnwidth]{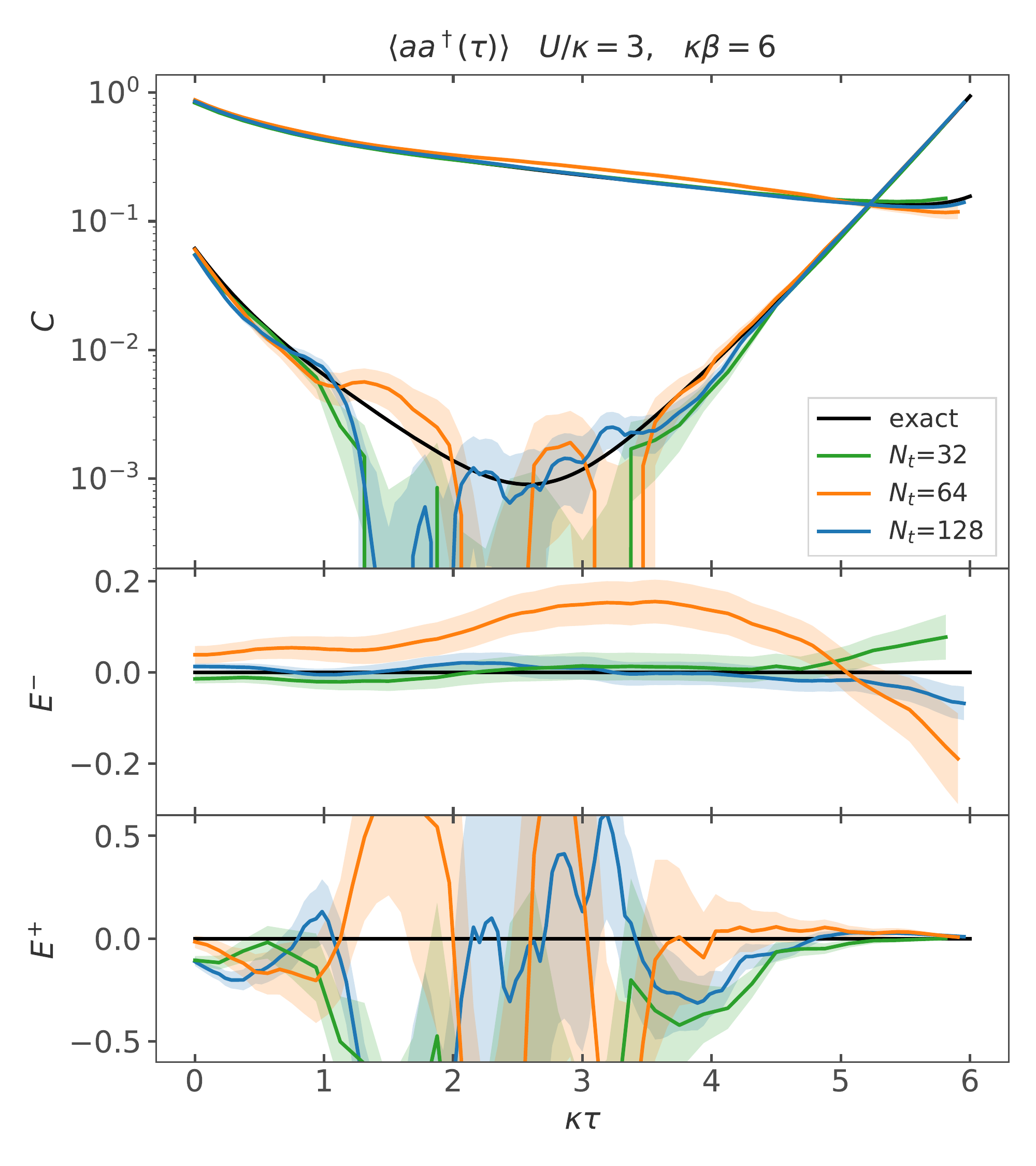}
  \includegraphics[width=.3\columnwidth]{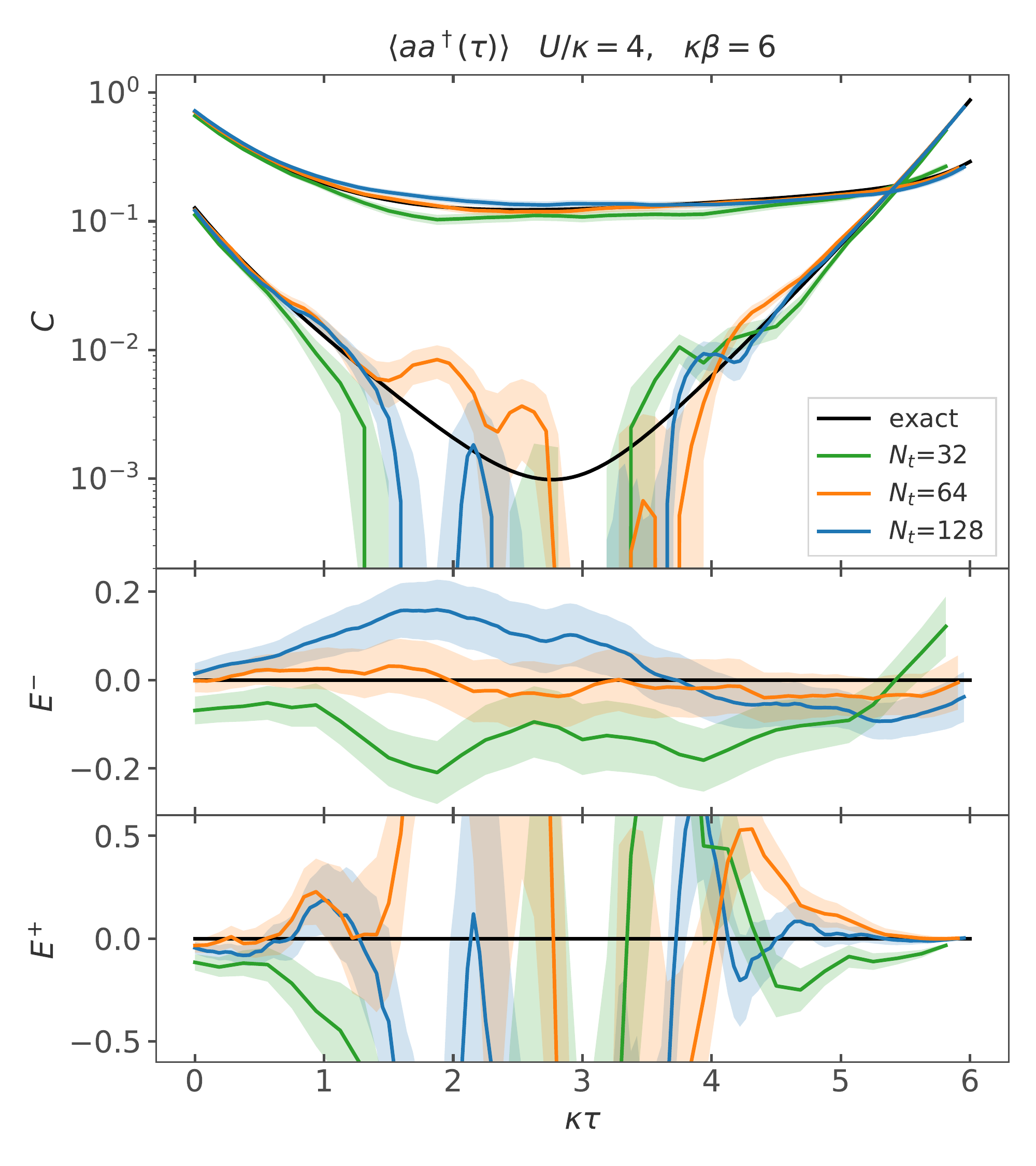}
  \includegraphics[width=.3\columnwidth]{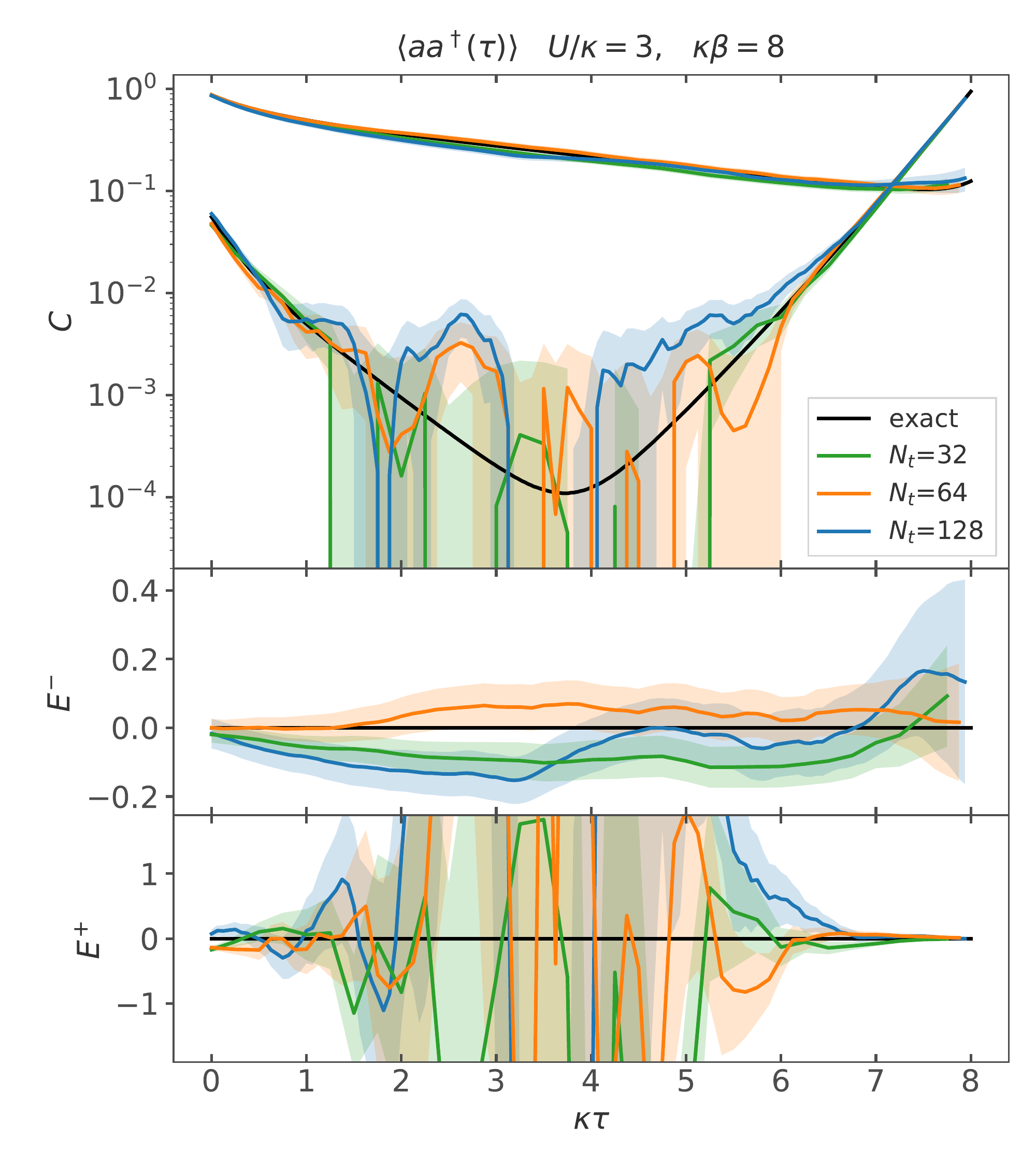}
  \caption{
    Single particle correlators $\langle a a^\dagger(\tau) \rangle$ on a triangle lattice for a sample size of $10^5$.
    The low energy correlators are averages of two degenerate results.
    The top panels show the correlators for different $N_t$ from HMC with a neural network and the results from exact diagonalization of the Hamiltonian.
    The lower panels show the relative error~\eqref{relative error}.
    Errors for the low and high energy correlators are labeled $E^-$ and $E^+$, respectively.\label{fig:triangle singleparticle corrs}}
\end{figure}

In \Figref{triangle singleparticle convergence} we show how the different methods' measurements of diagonalized\footnote{All correlators shown in this work are projected to irreps of the lattice, as in \eqref{triangle single particle operators}. We found this method suitable to produce diagonal all-to-all correlation matrices on triangle and tetrahedron lattices.} single-particle correlation functions converge as a function of the number of configurations.
With many configurations all three methods reproduce the exact correlators (shown in black).
However, the network reproduces the exact results (up to the expected $1/\sqrt{N}$ relative scale from statistical noise) with fewer configurations.

\begin{figure}[ht]
  \centering
  \includegraphics[width=.35\columnwidth]{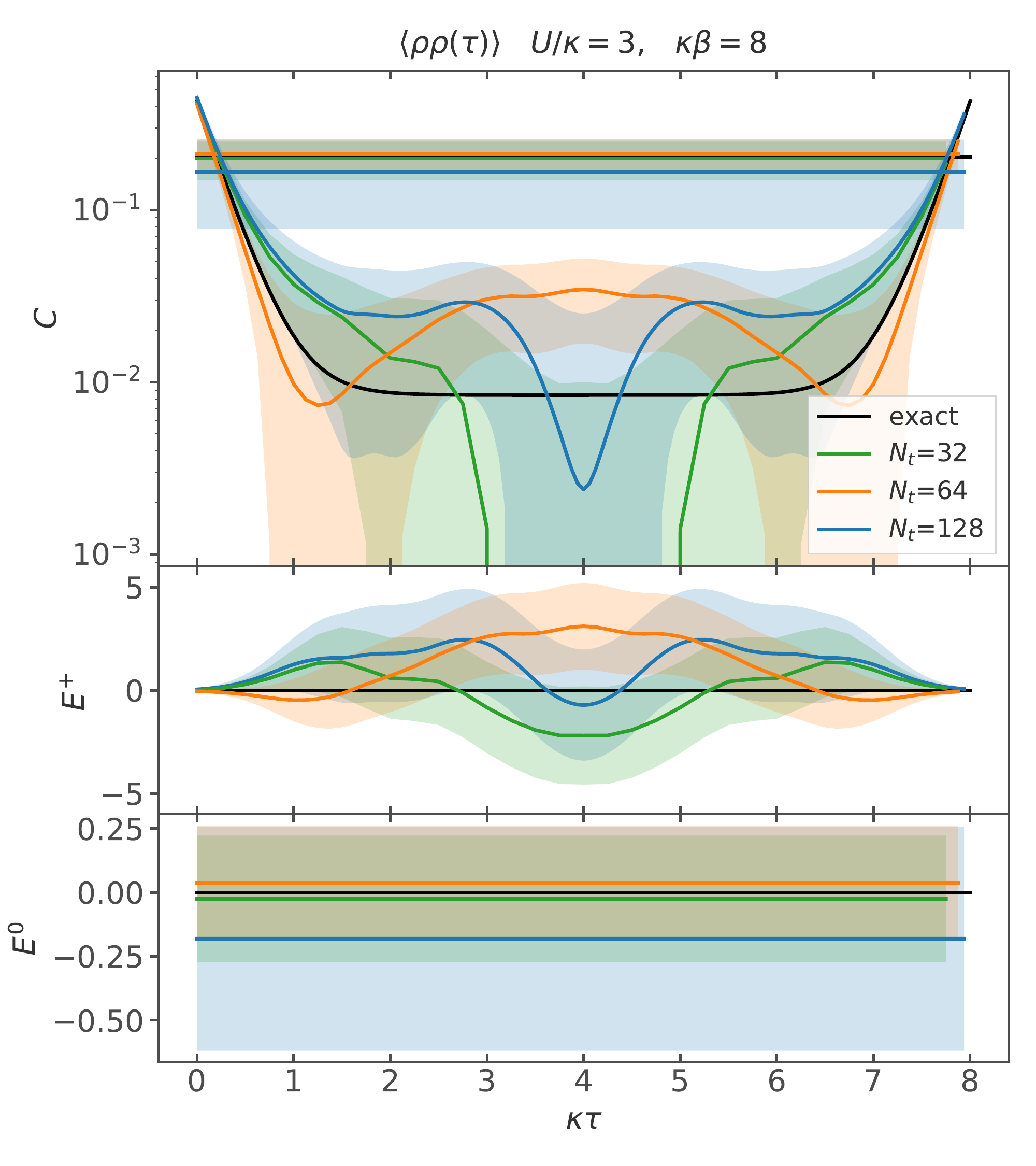}
  \includegraphics[width=.35\columnwidth]{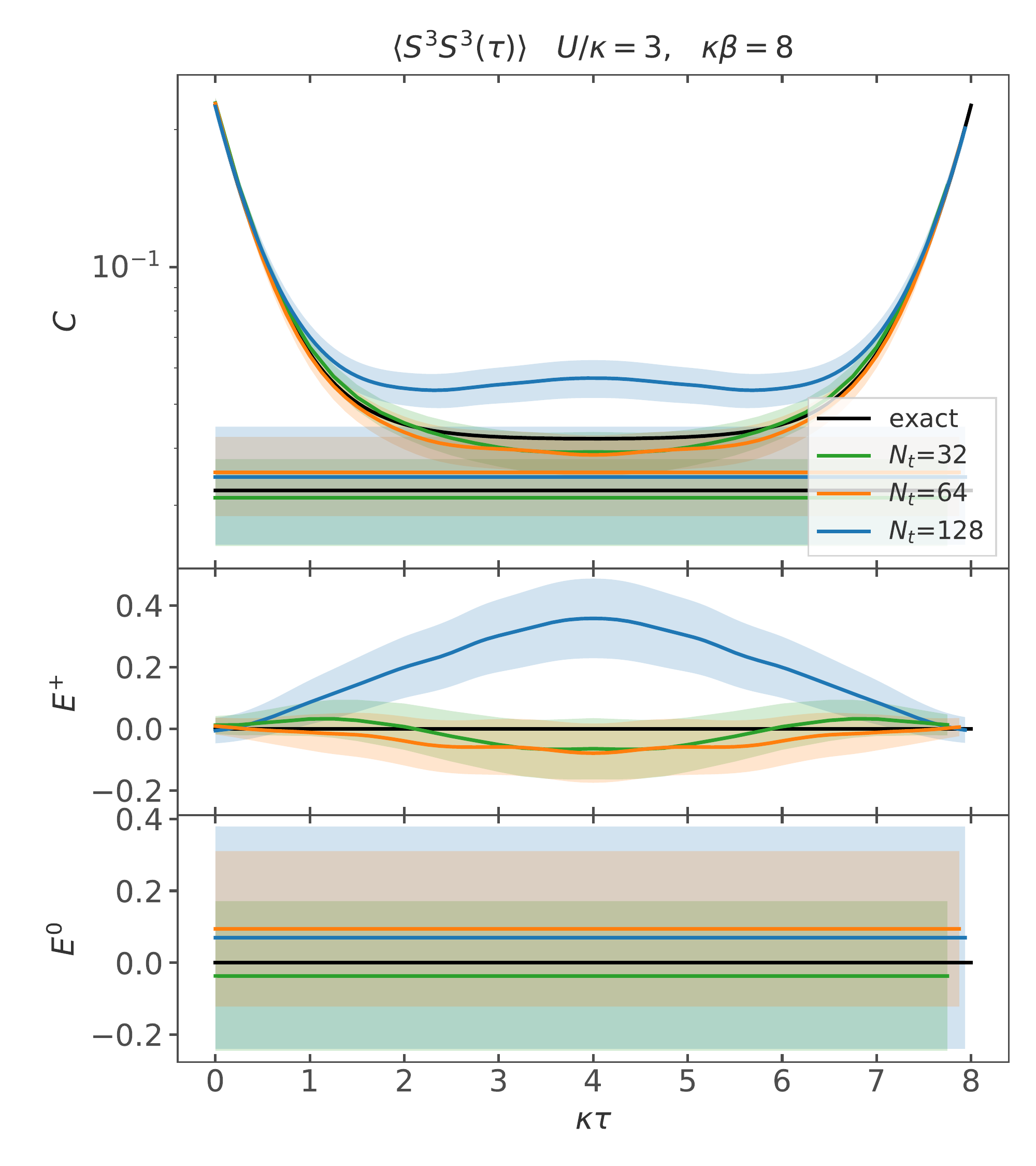}
  \caption{Correlation functions between two charge operators $\rho$ or two $S^3$ spin operators separated by euclidean time $\kappa\tau$ on a triangle for different discretization scales given by $\Nt$ and the exact result (in black).
    The bottom panels show the relative error~\eqref{relative error} for the constant correlator $(0)$ and the average of two heavy correlators $(+)$ which have $k=0$ and $\pm1$, respectively.\label{fig:triangle bilinear correlators}
    }
\end{figure}

Focusing on learnifold-enhanced HMC calculations, \Figref{triangle singleparticle corrs} shows single particle correlators for several different parameters and discretization scales measured on ensembles of $10^5$ configurations.
The figure also shows deviations from the exact results as
\begin{equation}\label{eq:relative error}
    E=C_\text{MC}/C_\text{exact}-1\ .
\end{equation}
In addition, we compute and diagonalize a variety of correlation functions between bilinear operators, as mentioned in \Secref{correlators} and detailed in \Appref{bilinear correlators}.
In \Figref{triangle bilinear correlators} we show two --- the diagonalized charge-charge and spin-spin correlators.
Additional correlators can be found in \Figref{triangle other corrs}.

\begin{figure}[ht]
  \centering
  \includegraphics[width=.45\columnwidth]{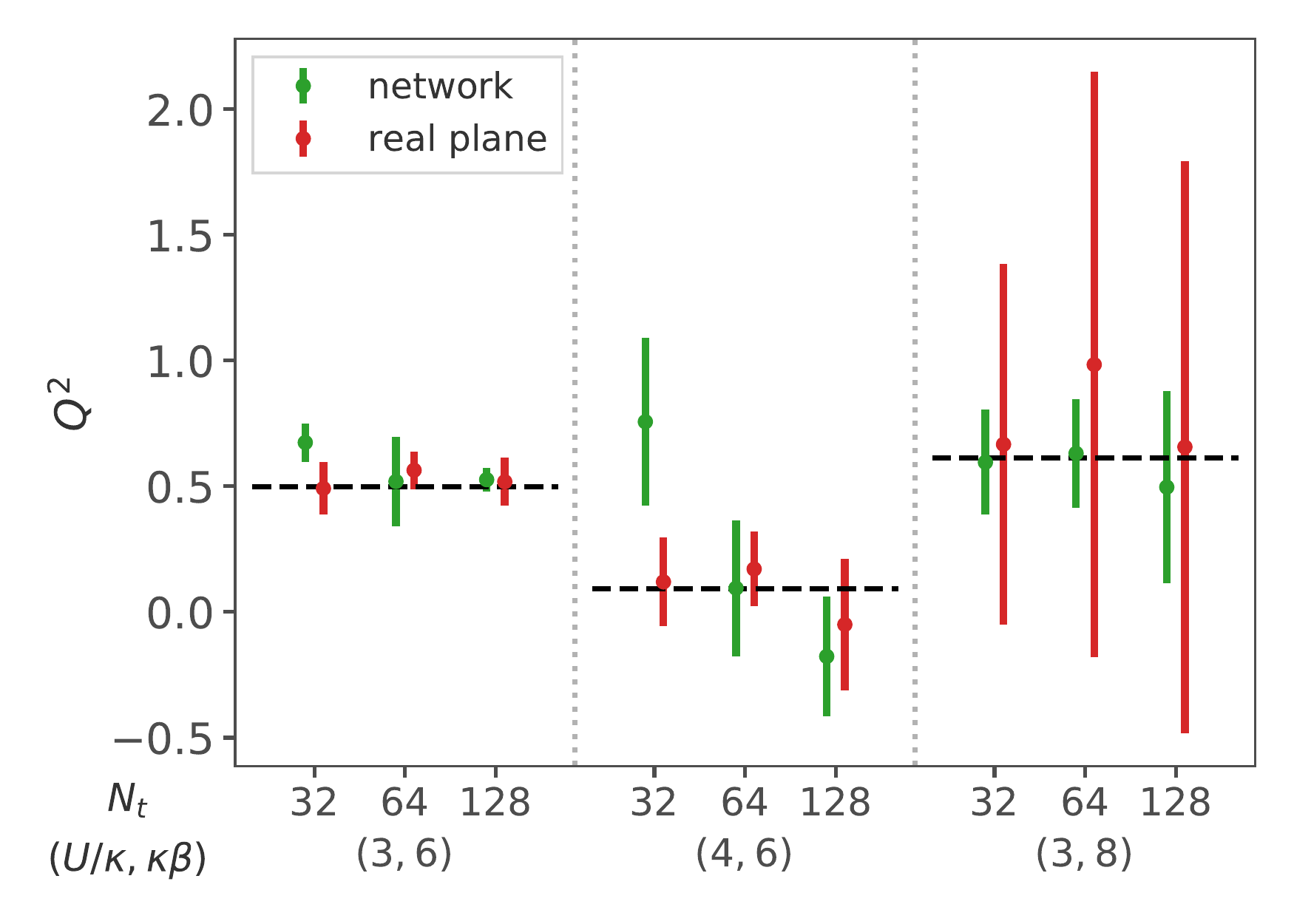}
  \includegraphics[width=.45\columnwidth]{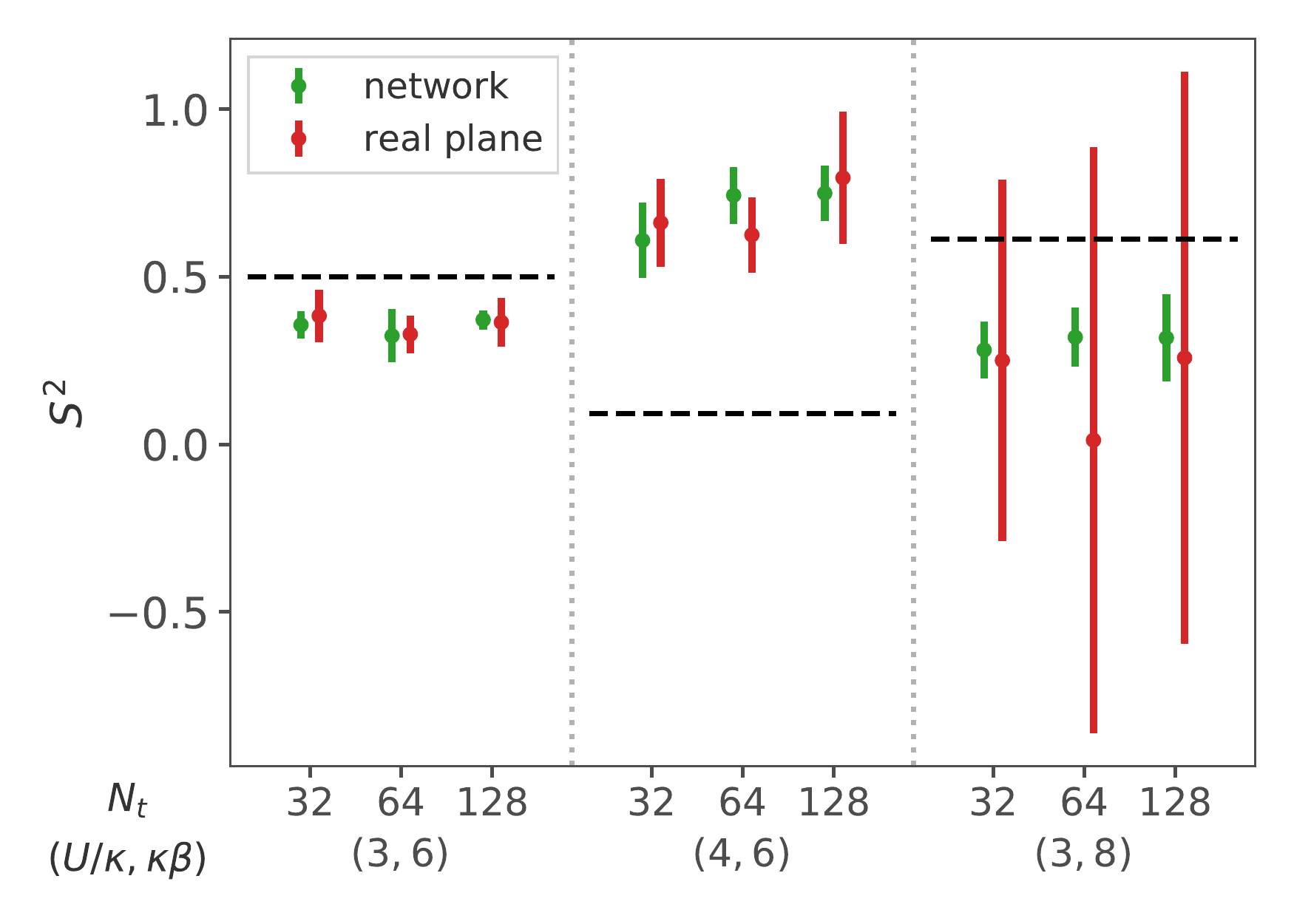}
  \caption{The order parameters $Q^2$ (left) and $S^2$ (right) calculated with the triangle system with various couplings $U$ and inverse temperatures $\beta$.  Shown is a comparison between results calculated with the neural network and on the standard real plane.\label{fig:triangle conserved quantities}}
\end{figure}

We can use the constant correlators to extract $\left\langle Q^2 \right\rangle$ and $\left\langle S^2 \right\rangle$, as explained in \Appref{conserved quantities}.
In \Figref{triangle conserved quantities} we show results for different couplings, temperatures, and discretizations.
The learnifold approach yields reduced errors and results consistent with the exact results of $Q^2$ for $U/\kappa = 3$.
The $(U/\kappa, \kappa\beta) = (4, 6)$ case seems to have a systematic deviation.
It is unsurprising that these parameters yield the worst result as they are closest to $U^\triangle/\kappa$.
The errors of $S^2$ are also reduced by the machine learning approach but a significant systematic deviation from the exact result remains in all cases.

\subsection{The Tetrahedron}\label{sec:tetrahedron}

Many arrangements of four sites are not bipartite.
Of these the ``most non-bipartite'', and therefore the one where we expect the worst sign problem, is the tetrahedron, where the connectivity matrix is proportional to the adjacency matrix of the complete graph on 4 sites,
\begin{equation}
    h = \kappa \left( \begin{array}{cccc}
		0	&	1	&	1	&   1   \\
		1	&	0	&	1	&   1   \\
		1	&	1	&	0   &   1   \\
        1   &   1   &   1   &   0   \\
	\end{array} \right)
\end{equation}
so that each subset of 3 sites forms a frustrated triangle; we label the sites 0-3.
The Hamiltonian commutes with the permutation operator $P$ that acts on any triangular face \eqref{triangle permutation} and leaves the other site alone.
We conventionally pick the symmetry axis through the fourth site to be the axis of symmetry around which we have a rotational quantum number.

The four irreducible single-particle destruction operators are
\begin{align}
    \label{eq:tetrahedron single particle operators}
    \hat{O}_{0}^0       &=  \frac{1}{2} \sum_{j=0}^3 a_j
        &
    \hat{O}_{1}^{0}     &=  \frac{1}{2\sqrt{3}}\left(3 a_3 - \sum_{j=0}^2 a_j\right)
        &
    \hat{O}_{1}^{\pm1}  &=  \frac{1}{\sqrt{3}} \sum_{j=0}^2 e^{\pm\frac{2\pi i}{3}j} a_j
\end{align}
where the lower index indicates an $L^2$-like quantum number $\ell$ and the upper index an $L_z$-like quantum number $m$.
The free Hamiltonian may be written in terms of these operators,
\begin{equation}
    H_0 = \kappa\left[
            \left(-3 \hat{O}_{0}^{0\dagger} \hat{O}_{0}^{0} + \sum_{m=-1}^{1} \hat{O}_{1}^{m\dagger} \hat{O}_{1}^{m}\right)
        -   (a \goesto b)
        \right]
\end{equation}
and the translationally-invariant interaction term transforms as an $\ell=0$, $m=0$ ``spherical tensor''.
Therefore, these are good quantum numbers and correlation functions put into this basis are diagonal.


\subsubsection{Results}\label{sec:tetrahedron results}

We expect the sign problem to be worse than on a triangle because the tetrahedron is substantially more frustrated since it has four triangular faces.
This is an opportunity to test our method in a system where calculations on the real plane are, as far as we can tell, just not possible.
\begin{figure}
    \includegraphics[width=.45\textwidth]{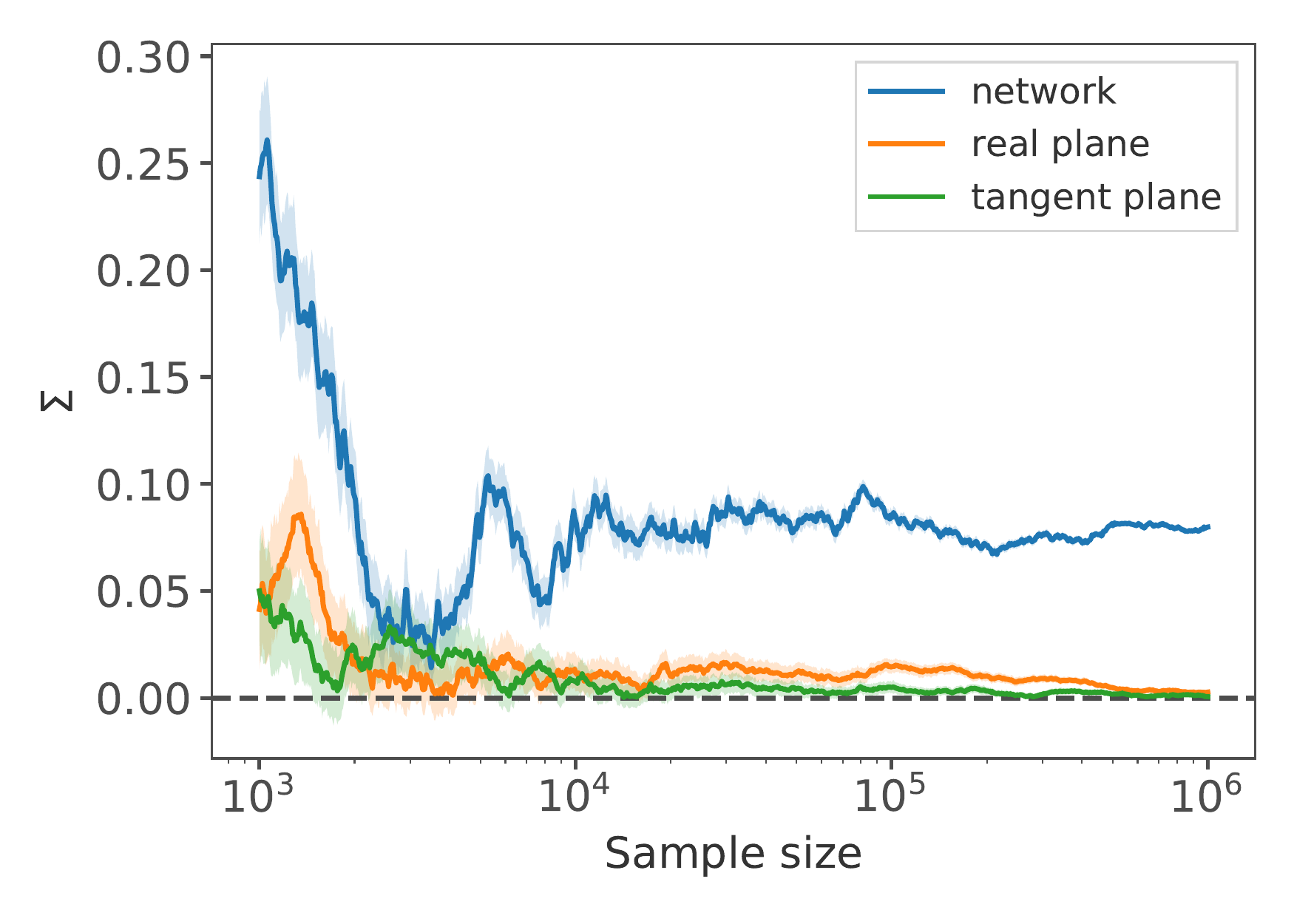} \includegraphics[width=.45\textwidth]{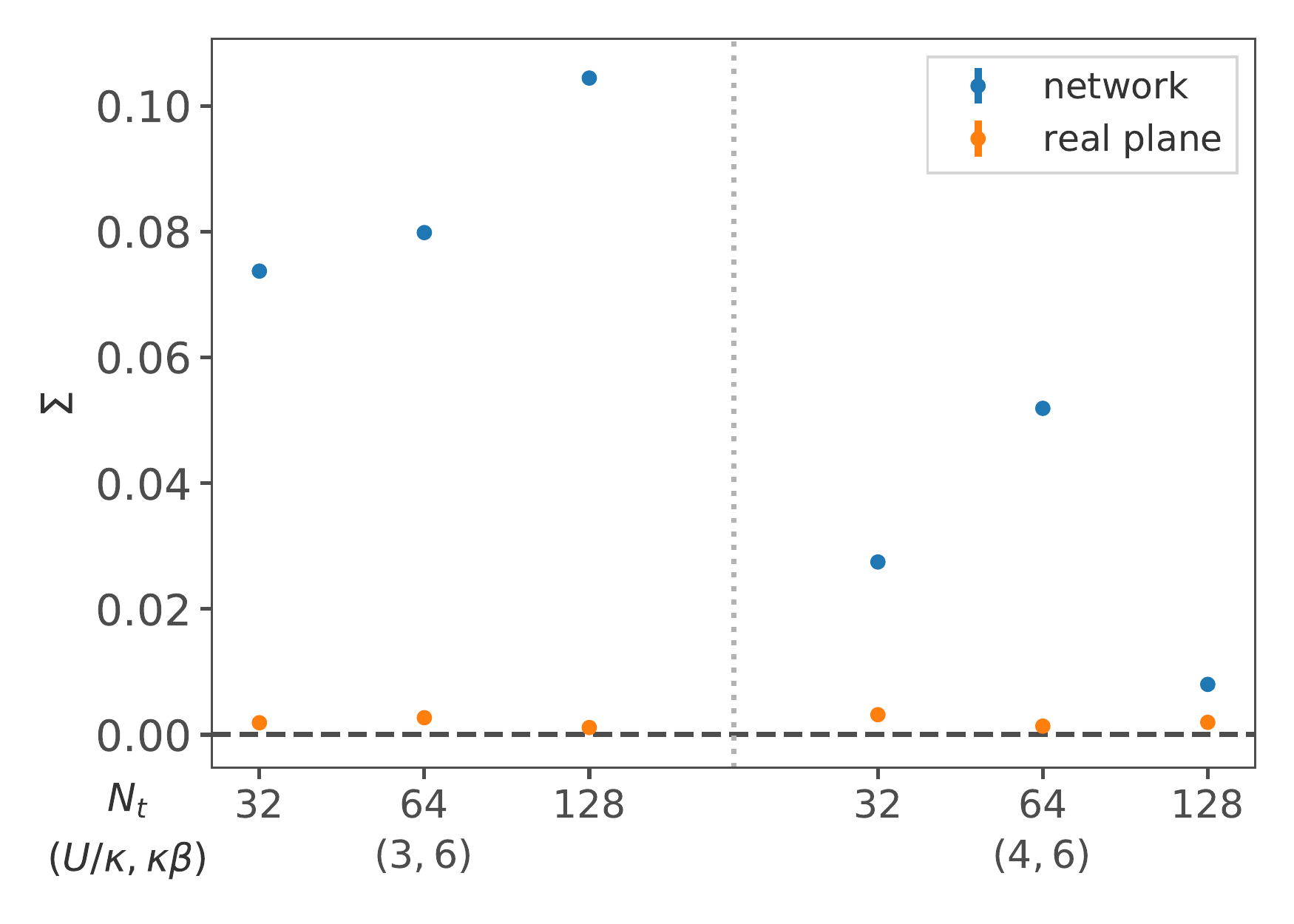}
    \caption{Left panel shows the statistical power as a function of sample size on a tetrahedron lattice with $N_t = 64$, $U/\kappa = 3$, $\kappa\beta = 6$.  Right panel shows statistical power for different parameters on a triangle lattice. Each point was estimated using $10^6$ configurations.\label{fig:tetrahedron statpower overview}}
\end{figure}
Indeed, HMC on both the real and tangent planes has essentially 0 statistical power as shown in the left panel of \Figref{tetrahedron statpower overview}, an extremely difficult sign problem.
In contrast, the neural network method converges to a finite statistical power.
Even though this value is small, it is sufficient as shown below.
The right panel of \Figref{tetrahedron statpower overview} shows this improvement in statistical power holds for all the ensembles we consider.

\begin{figure}[ht]
  \centering
  \includegraphics[width=0.9\columnwidth]{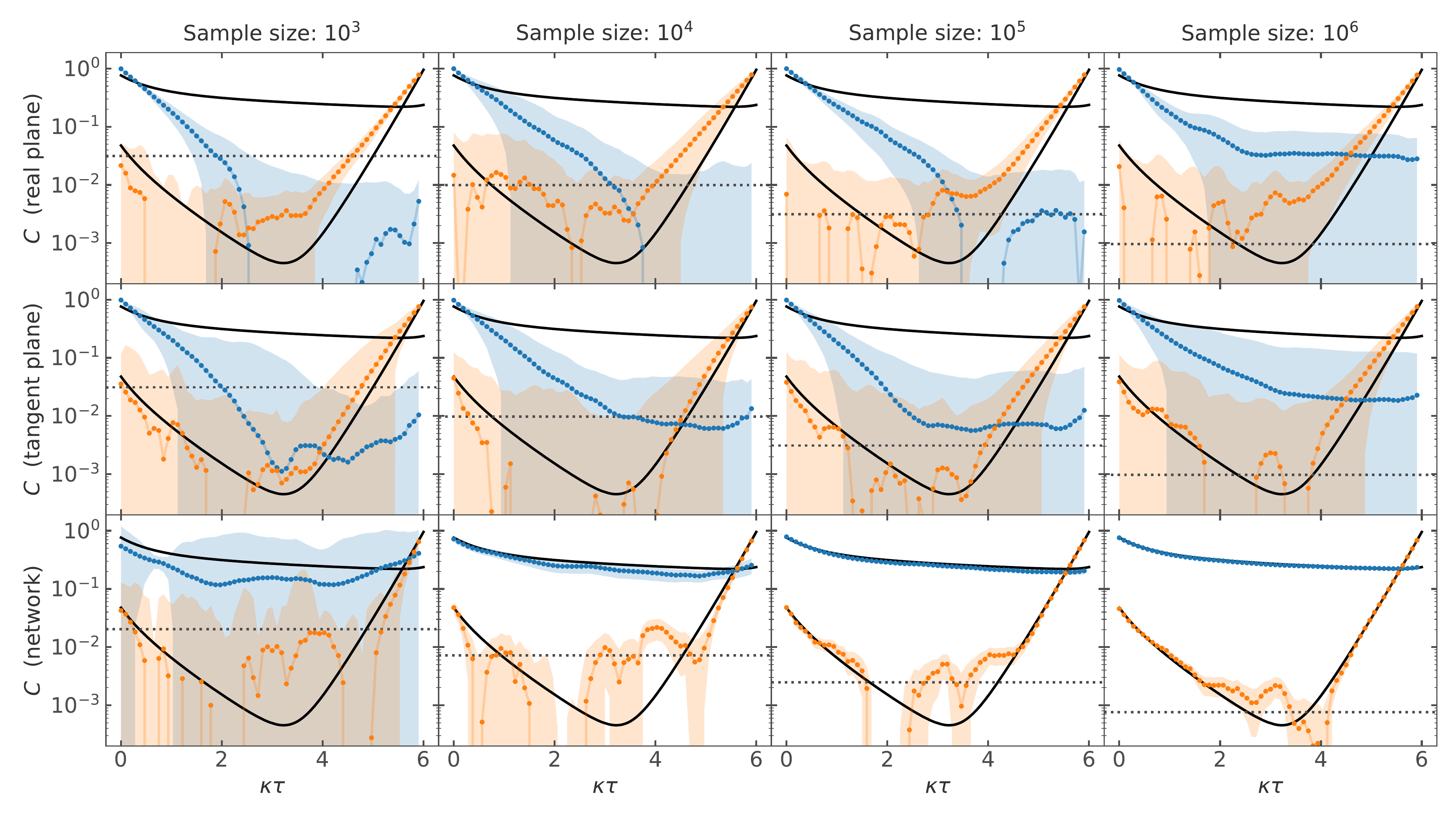}
  \caption{
    Single particle correlators $\langle a a^\dagger(\tau) \rangle$ on a tetrahedron lattice with $N_t = 64$, $U/\kappa = 3$, $\kappa\beta = 6$.
    Each column shows correlators obtained from ensembles of the given number of configurations, while each row shows a different implementation of HMC.
    Correlators are computed only on every $10^\text{th}$ configuration.
    The dotted lines are placed at $\max(C) / \sqrt{\text{Sample size}}$ and indicate the scale at which even a sign-problem-free method would show sizeable statistical fluctuations.\label{fig:tetrahedron singleparticle convergence}
  }
\end{figure}

\begin{figure}[ht]
  \centering
  \includegraphics[width=.37\columnwidth]{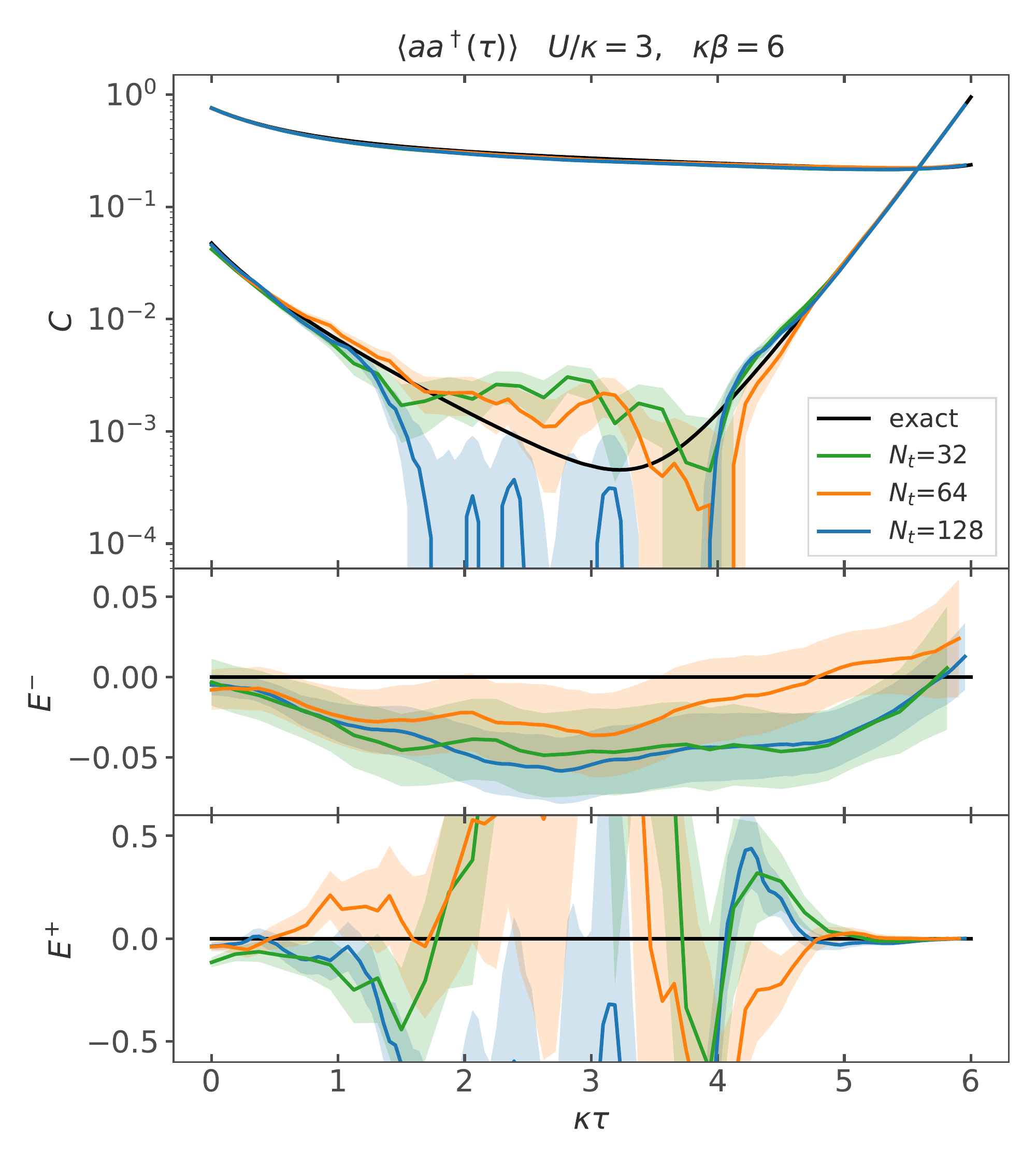}
  \includegraphics[width=.37\columnwidth]{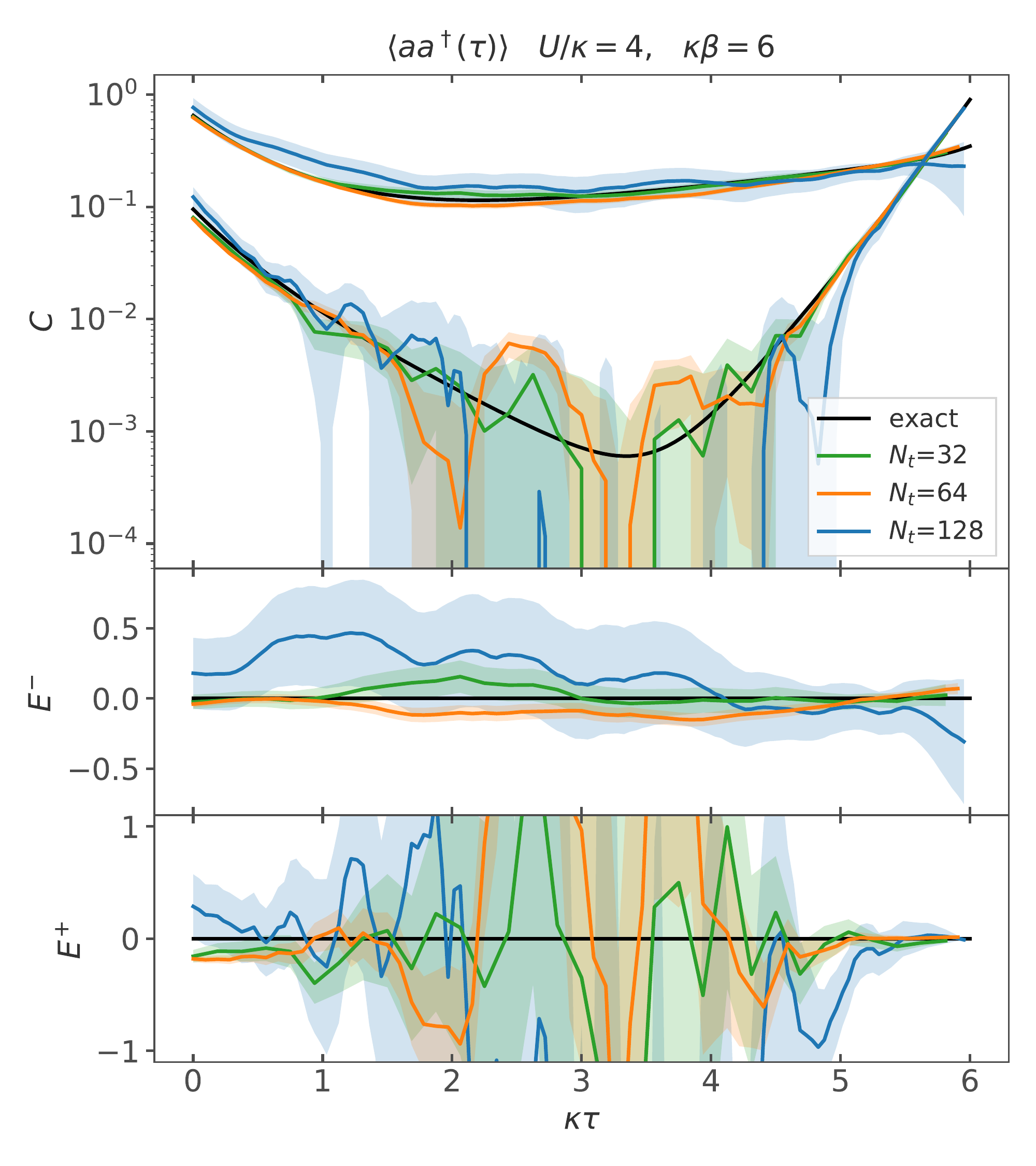}
  \caption{Single particle correlators $\langle a a^\dagger \rangle$ on a tetrahedron lattice.
    The low energy correlators are averages of three degenerate results.
    The top panels show the correlators for different $N_t$ from HMC with neural network and the results from exact diagonalization of the Hamiltonian.
    The lower panels show the relative error~\eqref{relative error}.
    Errors for the low and high energy correlators are labeled $E^-$ and $E^+$, respectively.\label{fig:tetrahedron singleparticle corrs}}
\end{figure}

\Figref{tetrahedron singleparticle convergence} shows the drastic improvement obtained when simulating with the neural network.
By studying the single-particle correlators, it is apparent that while the network method converges to the exact answer, the real and tangent plane methods are completely ineffective --- their uncertainties remain large and their match to the exact results poor.

In \Figref{tetrahedron singleparticle corrs} we show network-method results for different discretizations, and their relative errors $E$ \eqref{relative error}.
Those correlators were computed on an ensemble of $10^6$ configurations, measuring on only every $10^\text{th}$ configuration to reduce autocorrelations.
The light correlator is an average of the $\ell=1$ triplet of correlation functions.
The error of the heavy correlator grows for intermediate euclidean time, but this is expected given the concrete sample size, see \Figref{tetrahedron singleparticle convergence}.

\begin{figure}[ht]
  \centering
  \includegraphics[width=.4\columnwidth]{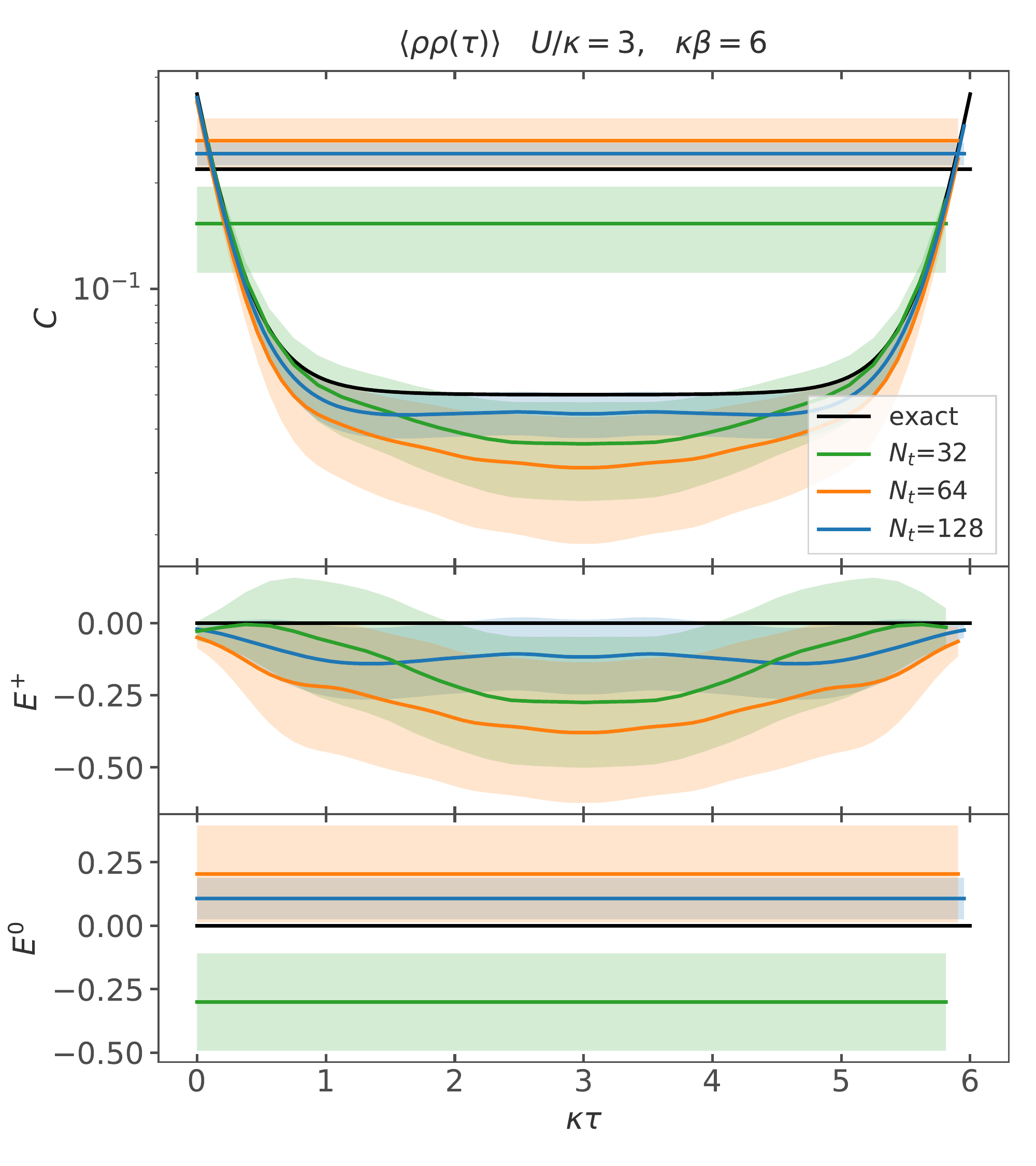}
  \includegraphics[width=.4\columnwidth]{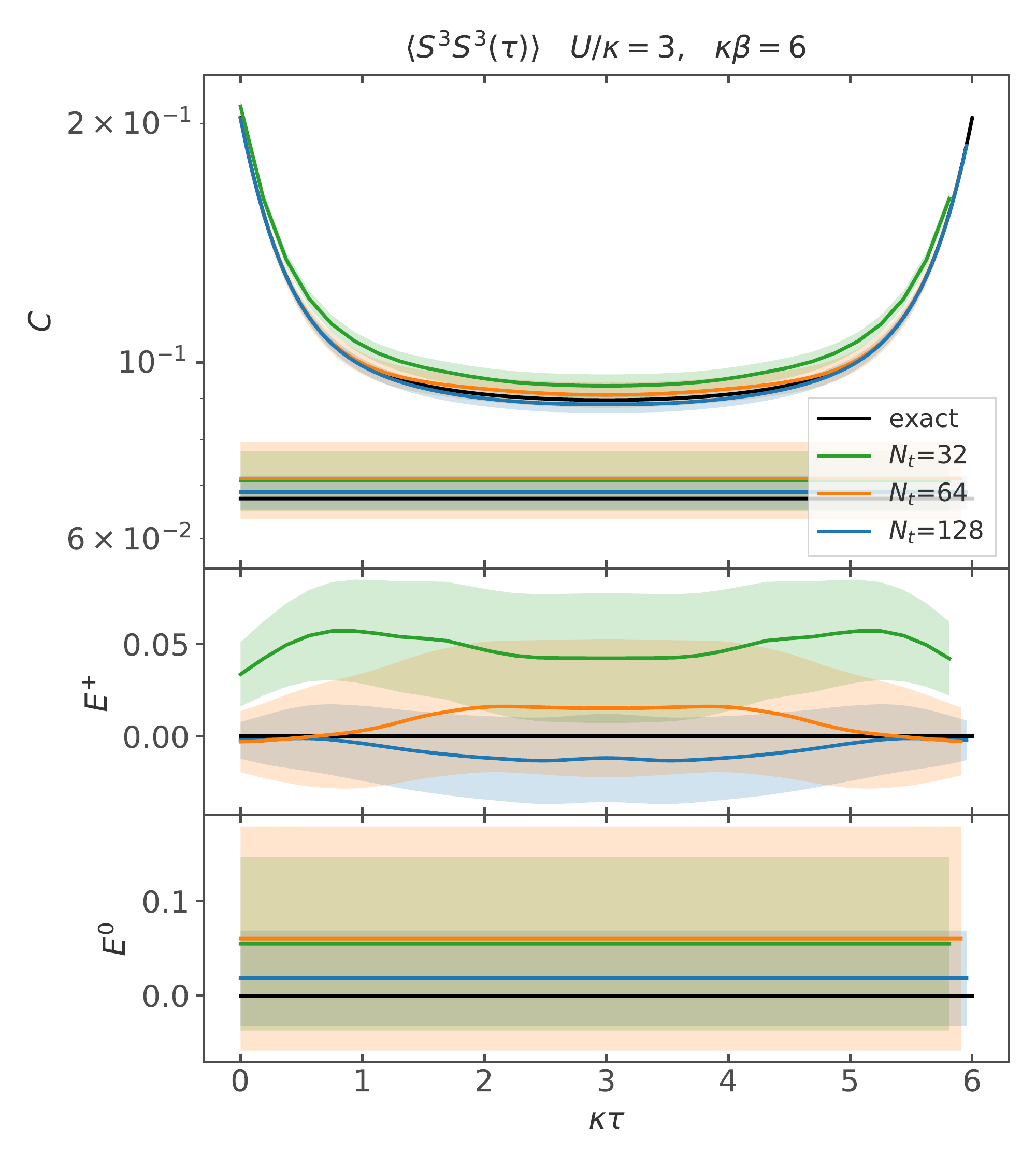}
  \caption{
    Correlation functions between two charge operators $\rho$ or two $S^3$ spin operators on a tetrahedron.
    The bottom panels show the relative error~\eqref{relative error} for the constant correlator $E^0$ and the average of three heavy correlators $E^+$, respectively.\label{fig:tetrahedron bilinear correlators}
    }
\end{figure}

In addition, we computed bilinear correlation functions as described in \Secref{correlators} and \Secref{bilinear correlators}.
\Figref{tetrahedron bilinear correlators} shows the continuum-limit convergence of the charge-charge and $S^3$-$S^3$ correlators projected to the singlet and triplet (as in~\eqref{tetrahedron single particle operators}) towards the exact result.
Additional examples can be found in \Figref{tetrahedron other corrs}.

\begin{figure}[ht]
  \centering
  \includegraphics[width=.45\columnwidth]{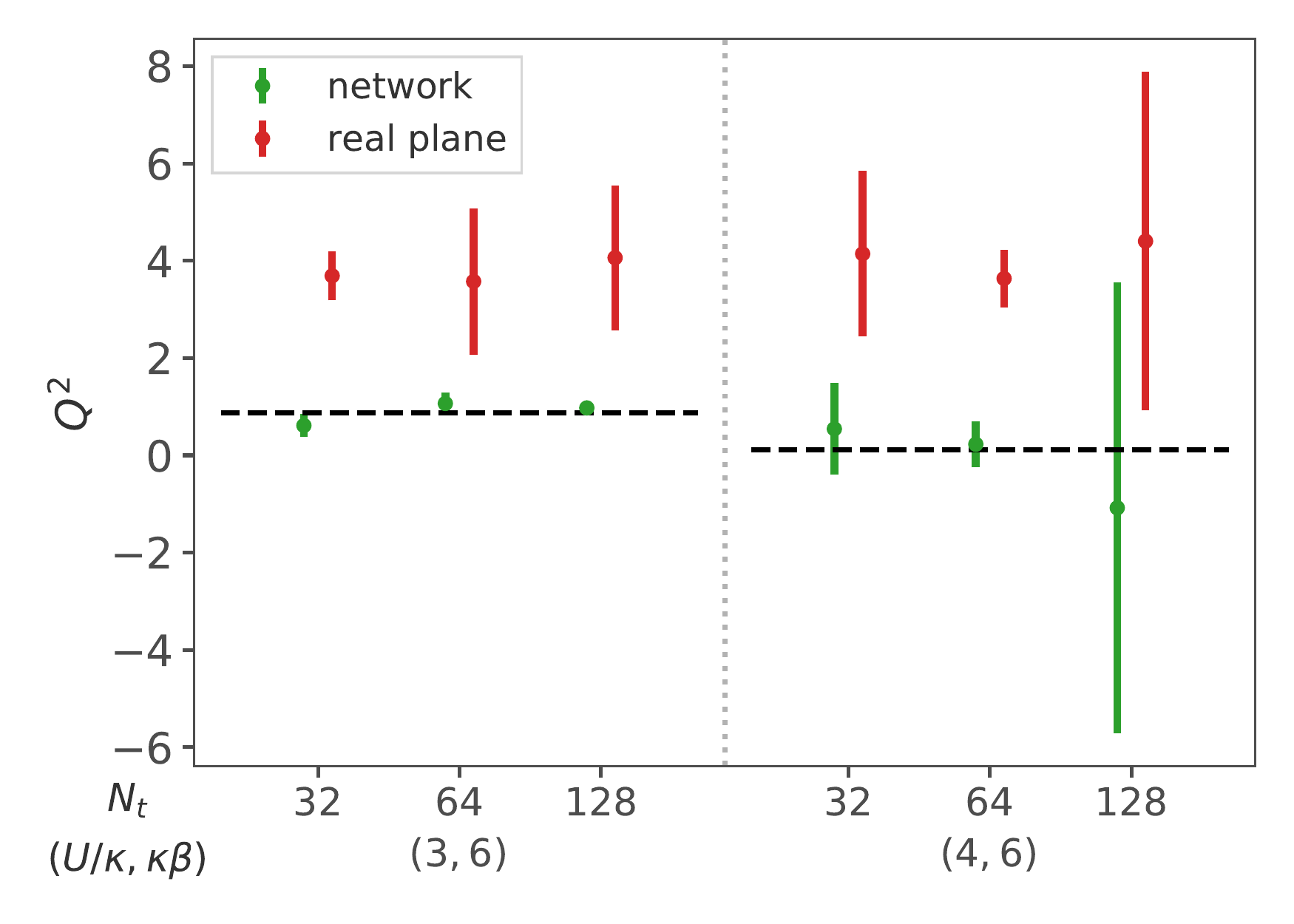}
  \includegraphics[width=.45\columnwidth]{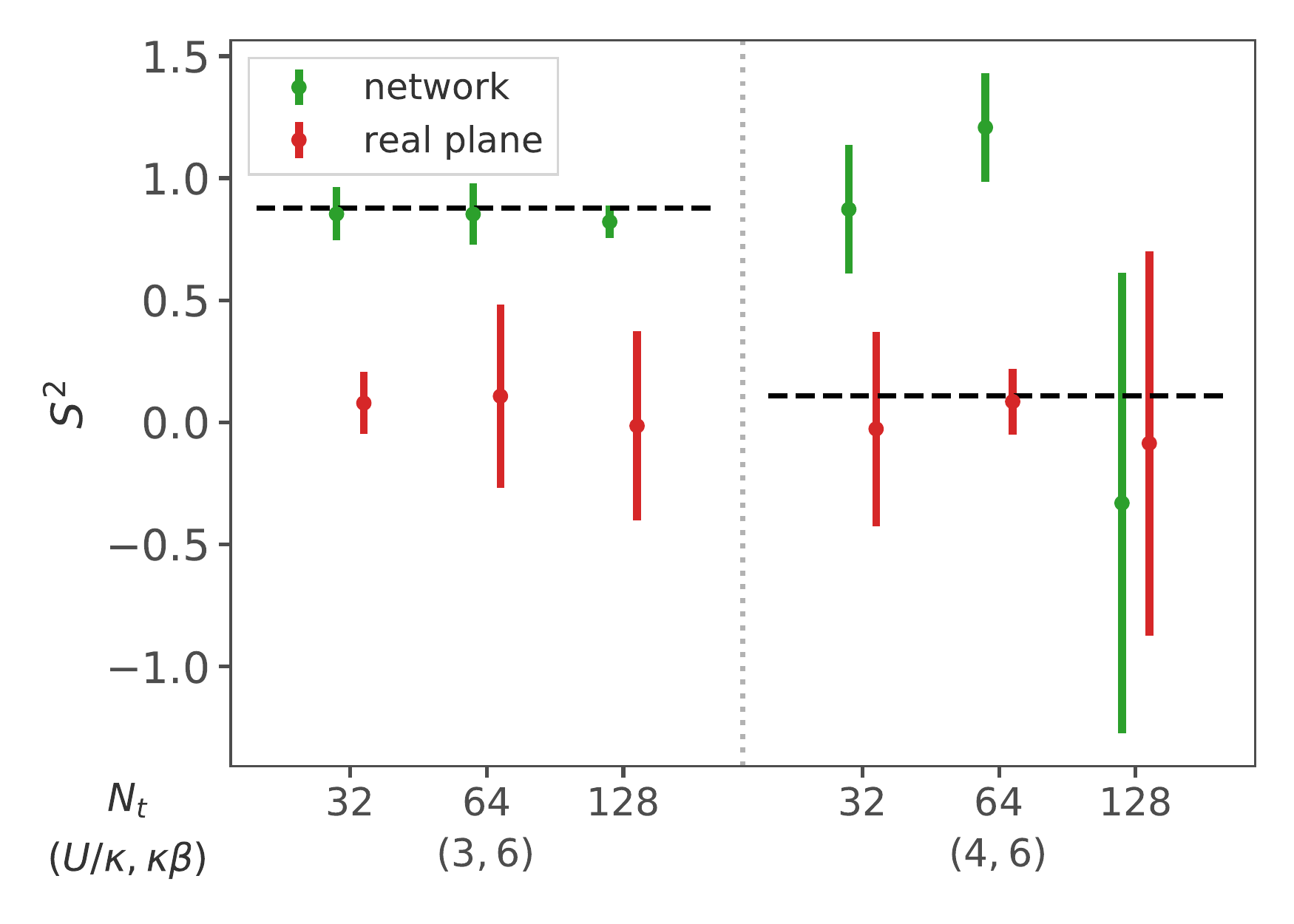}
  \caption{The order parameters $Q^2$ (left) and $S^2$ (right) calculated with the tetrahedron system with different couplings $U$ and inverse temperature $\beta$.  Shown is a comparison between results calculated with the neural network and on the standard real plane.\label{fig:tetrahedron conserved quantities}}
\end{figure}

In \Figref{tetrahedron conserved quantities} we show  $\left\langle Q^2 \right\rangle$ and $\left\langle S^2 \right\rangle$, see \Appref{conserved quantities} for their derivation.
Calculations on the real plane show significant systematic deviations from the exact result.
Calculations on learnifolds, however, have improved $Q^2$ for all $U/\kappa$ and $S^2$ for $U/\kappa = 3$ to the point where they agree with the exact result.
Curiously, $S^2$ for $(U/\kappa, \kappa\beta) = (4, 6)$ is worse on the learnifold.
Note, however, that while the real plane results are consistent with the exact value of $S^2$, they are also consistent with zero.

\section{Conclusions}\label{sec:conclusions}

In this work we adapted the learnifold method proposed by Alexandru, Bedaque, Lamm, and Lawrence~\cite{Alexandru:2017czx} to alleviate the sign problem in small, frustrated Hubbard model examples.
As shown in Figs.~\ref{fig:triangle statpower overview} and~\ref{fig:tetrahedron statpower overview}, our method definitively improves the statistical power, thus providing an exponential improvement in the sign problem.
With the NN, it reproduces results obtained from exact diagonalization on a variety of correlation functions, c.f. Figures \ref{fig:triangle singleparticle corrs}, \ref{fig:triangle bilinear correlators}, \ref{fig:triangle other corrs}, \ref{fig:tetrahedron singleparticle corrs}, \ref{fig:tetrahedron bilinear correlators}, \ref{fig:tetrahedron other corrs}.
The agreement with exact results provides \emph{a posteriori} evidence that we have an accurate, ergodic method.

By approximating the holomorphic flow with a neural network, we have developed an HMC method well-suited for tackling sign problems in the Hubbard model.
As opposed to most previous work involving Lefschetz thimbles, our method does not rely on obtaining the precise location of the main thimble nor on the locations of less critical thimbles. In fact, our neural networks perform better when not using prior knowledge of thimble locations.
The reason is two-fold.
First, during HMC evolution, the gaussian part of our action drives configurations to regions that are not dominated by a single thimble, bur rather multiple thimbles.
If we instead only trained on the main thimble, HMC would quickly force configurations into regions outside the trained network and we immediately encounter a numerical runaway problem --- the imaginary part of the fields become arbitrarily large in an attempt to minimize the action.
Second, our goal is to alleviate the sign problem to a point where standard reweighting techniques can be applied to obtain sufficiently accurate observables.
Thus our maximum flow time was relatively small, yet sufficient to obtain accurate results.
The modest flow times also prevented us from coming too close to points where the integration kernel vanishes, which in turn kept any ergodicity issues at bay.
Though integrating directly on the thimbles eliminates the sign problem altogether, we believe that determining the exact location of these manifolds and the subsequent manifold integration is a daunting task (perhaps as difficult as the original sign problem), and becomes only more difficult if one considers gauge theories in higher dimensions.

A drawback of our method is still the need to calculate the determinant of a Jacobian induced by the complexified fields.
Though our network was able to calculate the Jacobian much more quickly than a direct numerical flowing of the Jacobian as needed by holomorphic flow, the calculation still scales with the cubic power of the spacetime volume.
Thus we anticipate that calculations of larger dimensional systems with our network method will ultimately run into this scaling barrier.
We are actively researching methods to reduce this computational scaling, and we see some promise in networks that work directly with complex variables, see Appendix~\ref{sec:coupling layers}.

The sign problem we addressed was due to the non-bipartite spatial lattices, thus providing a basis for extensions to quasi-zero-dimensional systems like $C_{20}$ and the buckyball $C_{60}$\footnote{We do not anticipate the calculation of the determinant of the Jacobian to be too onerous for these systems.}.
Our framework is nonetheless easily adapted to other sign problems.
In future work, we anticipate simulating systems away from half filling and extended models with sufficiently strong non-local couplings~\cite{Buividovich:2016tgo}, both of which suffer a sign problem not only extensive in euclidean time but also in space.



\section*{Acknowledgements}
This work was done in part through financial support from the Deutsche Forschungsgemeinschaft (CRC 55 and the Sino-German CRC 110).
E.B. is supported by the U.S. Department of Energy under Contract No. DE-FG02-93ER-40762.
The authors gratefully acknowledge the computing time granted through JARA-HPC on the supercomputer JURECA~\cite{jureca} at Forschungszentrum J\"ulich. 


\appendix
\section{Correlation Functions of Bilinear Operators}\label{sec:bilinears}


Spin-spin correlation functions are correlation functions between local spin operators $S_x^i$ where $x$ is a lattice site and $i$ runs over the indices of the Pauli matrices,
\begin{equation}
    S_x^i = \frac{1}{2} \sum_{ss'} c_{xs} \sigma^i_{ss'} \adjoint{c}_{xs'}
\end{equation}
and where $c$ is a doublet of operators,
\begin{equation}
    c_{xs} = \left(\begin{array}{c} a_x \\ (-\sigma_\kappa)^x \adjoint{b}_x \end{array}\right)
\end{equation}
where $\sigma_\kappa$ is $+1$ on bipartite lattices and must be $-1$ on non-bipartite lattices, following the convention of Isle (bipartite graphs can also have $\sigma_\kappa=-1$).
In \Ref{Buividovich:2018yar} the authors also define, just after (1), the electric charge operator
\begin{equation}
    \rho_x
    =   \adjoint{c}_{x,\uparrow}   c_{x,\uparrow}
    +   \adjoint{c}_{x,\downarrow} c_{x,\downarrow}
    -   1
\end{equation}
which can be rewritten as $ \rho_x = 1-2 S^0_x$, where the $0^{th}$ Pauli matrix is the $2\times2$ identity matrix.  The $S$ operators are Hermitian.
Rewriting those operators into the Isle basis,
\begin{align}
    \label{eq:spin bilinears}
    S^0_x &= \frac{1}{2} \left[ a_x \adjoint{a}_x - b_x \adjoint{b}_x +1 \right]
    &
    S^1_x &= \frac{1}{2} (-\sigma_\kappa)^x \left[ \adjoint{b}_x \adjoint{a}_x + a_x b_x \right]
    \nonumber\\
    \rho_x &= n^a_x - n^b_x = 1-2S^0_x = b_x\adjoint{b}_x - a_x \adjoint{a}_x
    &
    S^2_x &= \frac{i}{2} (-\sigma_\kappa)^x \left[ \adjoint{b}_x \adjoint{a}_x - a_x b_x \right]
    \\
    n_x &= n^a_x + n^b_x = 1-2S^3_x = a^\dagger_x a_x + b^\dagger_x b_x
    &
    S^3_x &= \frac{1}{2} \left[ a_x \adjoint{a}_x + b_x \adjoint{b}_x -1 \right]
    \nonumber
\end{align}
where the $\sigma_\kappa$ squares away when two $b$ operators are multiplied and no sum is implied on the right-hand sides.
The other two spin bilinears have absolute charge 2,
\begin{align}
    {(S\pp)}_x &= \adjoint{a}_x b_x
    &
    {(S\mm)}_x &= \adjoint{b}_x a_x.
\end{align}
There are, of course, other doubly-charged operators but none that live on a single site, by Pauli exclusion.

The three spin operators obey the commutation relation
\begin{equation}
    \label{eq:spin algebra}
    \left[ S_x^i, S_y^j \right] = i \delta_{xy} \epsilon^{ijk} S_x^k
\end{equation}
which may be checked explicitly by writing out the operators and using the anticommutation properties of $a$ and $b$.
By a similar exercise one may show
\begin{equation}
    \label{eq:charge density commutators}
    \left[ S_x^0, S_y^j \right] = 0.
\end{equation}

Single-particle and single-hole operators $\mathcal{O}$ with a definite third component of spin $s_3$ obey the operator eigenvalue equation
\begin{equation}
    \label{eq:S3 eigenoperator equation}
    [ S_x^3, \mathcal{O}_y ] = s_3 \mathcal{O}_y \delta_{xy}.
\end{equation}
This equation is satisfied when
$(\mathcal{O},s_3) = (a, +\frac{1}{2})$,
$(\adjoint{a},           -\frac{1}{2})$,
$(b,                     +\frac{1}{2})$, and
$(\adjoint{b},           -\frac{1}{2})$.
Single-particle and single-hole operators $\mathcal{O}$ with a definite electric charge $q$ obeys the operator eigenvalue equation
\begin{equation}
    \label{eq:rho eigenoperator equation}
    [ \rho_x, \mathcal{O}_y ] = q \mathcal{O}_y \delta_{xy}.
\end{equation}
This equation is satisfied when
$(\mathcal{O},q) = (a,          -1)$,
                  $(\adjoint{a},+1)$,
                  $(b,          +1)$, and
                  $(\adjoint{b},-1)$.
Note that the signs differ from the $S^3$ case.

One may also construct spin raising and lowering operators in the standard way,
\begin{align}
    S^+_x = S^1_x + i S^2_x &= (-\sigma_\kappa)^x a_x b_x
    &
    S^-_x = S^1_x - i S^2_x &= (-\sigma_\kappa)^x \adjoint{b}_x \adjoint{a}_x
    \label{eq:spin ladders}
\end{align}
which obey the eigenvalue relations
\begin{equation}
    [S_x^3, S^\pm_y ] = \pm S^\pm_y \delta_{xy},
\end{equation}
which can be shown using the single-particle and single-hole eigenvalue equations and the Leibniz rule.

The construction of the number operators proceeds in a similar fashion,
\begin{align}
    \delta_{xx} - n^p_x &= S^0_x + S^3_x = a_x \adjoint{a}_x = \delta_{xx} - \adjoint{a}_x a_x
    \\
    n^h_x &= S^0_x - S^3_x = -b_x \adjoint{b}_x + \delta_{xx} = -\delta_{xx} + \adjoint{b}_x b_x + \delta_{xx} = \adjoint{b}_x b_x.
    \label{eq:number operators}
\end{align}
We can of course drop the constant term in the first definition.  The number operators obey the equations
\begin{align}
    \left[ n^p_x, a_y\right] &= - a_y \delta_{xy}
    &
    \left[ n^p_x, \adjoint{a}\right] &= + \adjoint{a}_y \delta{xy}
\end{align}
and similarly for holes.
It is easy to see using the eigenoperator equations \eqref{S3 eigenoperator equation} and \eqref{rho eigenoperator equation} and the Leibniz rule that these operators commute with the local electric charge and spin, so that they have vacuum quantum numbers, while the doubly-charged operators satisfy
\begin{align}
    [\rho_x, {(S\pp)}_y] &= +2 {(S\pp)}_y \delta_xy
    &
    [\rho_x, {(S\mm)}_y] &= -2 {(S\mm)}_y \delta_xy
\end{align}
and are spin-0 because they commute with the spin operators.
The one-point functions may be computed by Wick contraction
\begin{align}
    \frac{1}{N_t}\sum_t \langle n^p_x \rangle
    &=
    \left\langle 1- \frac{1}{N_t}\sum_t P_{xtxt} \right\rangle
    &
    \frac{1}{N_t}\sum_t \langle n^h_x \rangle
    &=
    \left\langle 1- \frac{1}{N_t}\sum_t H_{xtxt} \right\rangle
    \label{eq:one point functions}
\end{align}
where we denoted the Wick contraction of $a_{x,t_f} \adjoint{a}_{y,t_i} = (M^p)^{-1}_{xt_fyt_i} \equiv P_{xt_fyt_i}$, defining the particle propagator $P$,
and similarly for holes $b_{x,t_f} \adjoint{b}_{y,t_i} = (M^h)^{-1}_{xt_fyt_i} \equiv H_{xt_fyt_i}$ the hole propagator.
These may be combined according to \eqref{spin bilinears} to get one-point expectation values for $n_x$, $\rho_x$, and $S^3_x$.
Bilinears not having vacuum quantum numbers have vanishing one-point expectation values.


\subsection{Correlation Functions}\label{sec:bilinear correlators}

Now we can write two-point correlation functions
\begin{equation}
    C^{uv}_{xy}(\tau) = \frac{1}{N_t} \sum_t \left\langle S^{u}_{x,t+\tau} S^{v}_{y,t} \right\rangle
\end{equation}
and we do not need to track time separately, until we start analyzing how to actually analyze these correlation functions via their spectral decompositions, though we always put the $x$ position at the initial time $i=t$ and the $y$ position at the final time $f=t+\tau$, so one can read $x$ and $y$ as superindices.
When calculating numerically we sum over all initial timeslices $t$ to ensure the only time dependence is on the time difference $\tau$.
\emph{ In these correlator expressions $x$ and $y$ are unsummed.}

The simplest correlation function is between $S^+$ and $S^-$,
\begin{align}
    C^{+-}_{xy} =
    \left\langle S^+_x S^-_y \right\rangle
        = (-\sigma_\kappa)^{x+y}\left\langle a_x b_x \adjoint{b}_y \adjoint{a}_y \right\rangle
        &= (-\sigma_\kappa)^{x+y}\left\langle P_{xy} H_{xy}\right\rangle
    \label{eq:C+-}\\
    C^{-+}_{xy} =
    \left\langle S^-_x S^+_y \right\rangle
        = (-\sigma_\kappa)^{x+y}\left\langle \adjoint{b}_x \adjoint{a}_x a_y b_y  \right\rangle
        &= (-\sigma_\kappa)^{x+y}\left\langle (\delta_{yx}- b_y \adjoint{b}_x)(\delta_{yx} - a_y \adjoint{a}_x)  \right\rangle
        \nonumber\\
        &= (-\sigma_\kappa)^{x+y}\left\langle (\delta_{yx}- H_{yx})(\delta_{yx} - P_{yx})  \right\rangle
    \label{eq:C-+}
\end{align}
where we have taken advantage of the anticommutator rules and that the Wick contractions yield the particle and hole propagators $P$ and $H$ (suppressing the time dependence for clarity).
At half filling on a bipartite lattice, the cost to create or destroy a spin from the vacuum should be equal and the correlators should match, in the limit of large statistics.
At equal time $\tau=0$, these correlation functions provide access to the spin-flip information~\cite{10.21468/SciPostPhys.7.5.064}.

Correlations between the number operator $n^p$ and itself or $n^h$ are also simple to write,
\begin{align}
    C^{ph}_{xy} =
    \left\langle n^p_x n^h_y \right\rangle
        =  \left\langle (\delta_{xx}-a_x \adjoint{a}_x)(\delta_{yy}-b_y\adjoint{b}_y) \right\rangle
        &=  \left\langle (\delta_{xx}-P_{xx})(\delta_{yy}-H_{yy}) \right\rangle
    \label{eq:particle-hole correlator}
    \\
    C^{pp}_{xy} =
    \left\langle n^p_x n^p_y \right\rangle
        =  \left\langle (\delta_{xx}-a_x \adjoint{a}_x)(\delta_{yy}-a_y\adjoint{a}_y) \right\rangle
        &=  \left\langle  \delta_{xx}\delta_{yy}
                        - a_x\adjoint{a}_x \delta_{yy}
                        - \delta{xx} a_y\adjoint{a}_y
                        + a_x\adjoint{a}_xa_y\adjoint{a}_y \right\rangle
    \nonumber\\
        &=  \left\langle \delta_{xx}\delta_{yy}
                        - a_x\adjoint{a}_x \delta_{yy}
                        - \delta_{xx}a_y\adjoint{a}_y
                        + a_x(\delta_{xy}-a_y \adjoint{a}_x)\adjoint{a}_y \right\rangle
    \nonumber\\
        &= \left\langle   \delta_{xx}\delta_{yy}
                        - P_{xx} \delta_{yy}
                        - \delta_{xx} P_{yy}
                        + P_{xy}\delta_{xy}
                        + P_{xx}P_{yy}
                        - P_{xy}P_{yx} \right\rangle
    \label{eq:particle-particle correlator}
\end{align}
and we can interchange the p/h species superscripts by exchanging the $P$ and $H$ propagators.%
While these correlators are between operators as simple as $S^+$ and $S^-$, computationally these Wick contractions are tougher to compute because they are ``quark-line disconnected''.

We can also build correlators between the spin operators $S^i$.  For example $C^{11}$ is given by
\begin{align}
    C^{11}_{xy} =
    \left\langle S^{1}_{x} S^{1}_{y} \right\rangle
    &= \frac{1}{4} (-\sigma_\kappa)^{x+y} \left\langle \left[ \adjoint{b}_x \adjoint{a}_x + a_x b_x \right] \left[ \adjoint{b}_y \adjoint{a}_y + a_y b_y \right] \right\rangle \nonumber\\
        &= \frac{1}{4} (-\sigma_\kappa)^{x+y} \left\langle a_x \adjoint{a}_y b_x \adjoint{b}_y + (\delta_{yx} - a_y \adjoint{a}_x)(\delta_{yx} - b_y \adjoint{b}_x) \right\rangle \nonumber\\
    &= \frac{1}{4} (-\sigma_\kappa)^{x+y} \left\langle P_{xy} H_{xy} + (\delta_{yx} - P_{yx})(\delta_{yx} - H_{yx}) \right\rangle
    \label{eq:C11}
\end{align}
where we have used the fact that we will only get a non-zero result if we have the same number of $a$s as $\adjoint{a}$s (and likewise for $b$) to drop the four-dagger and no-dagger terms.
Computing $C^{22}_{xy}$ requires
\begin{align}
    \left\langle S^{2}_{x} S^{2}_{y} \right\rangle
    &= \frac{1}{4} (-\sigma_\kappa)^{x+y} \left\langle \left[ \adjoint{b}_x \adjoint{a}_x - a_x b_x \right] \left[ a_y b_y - \adjoint{b}_y \adjoint{a}_y \right] \right\rangle
\end{align}
though when written out in their complete glory, only the non-vanishing operator content in $\left\langle S^{1}_{x} S^{1}_{y}\right\rangle$ remains, so $C^{22}_{xy} = C^{11}_{xy}$ configuration-by-configuration (the vanishing operators have the opposite sign).
In fact, using the definition of the spin raising and lowering operators \eqref{spin ladders} one concludes
\begin{align}
    C^{11}_{xy} + C^{22}_{xy} &= \half          \left(C^{+-}_{xy} + C^{-+}_{xy}\right)
    &
    C^{12}_{xy} - C^{21}_{xy} &= \frac{i}{2}    \left(C^{+-}_{xy} - C^{-+}_{xy}\right)
\end{align}
and we have explicitly checked the first identity by computing the Wick contractions \eqref{C+-}, \eqref{C-+}, and \eqref{C11}, and the fact that $C^{22}_{xy} = C^{11}_{xy}$.
It is easy to show that once the all-dagger or no-dagger operators are dropped,
\begin{align}
    \left\langle S^1_x S^2_y \right\rangle &= - \left\langle S^2_x S^1_y \right\rangle
    &
    &\text{so that}&
    C^{12}_{xy} &= \frac{i}{4}\left(C^{+-}_{xy} - C^{-+}_{xy}\right)
\end{align}
which may be checked explicitly, and is true configuration-by-configuration.

The other two spins $S^0$ and $S^3$ do not enjoy such simplifications, because unlike the raising and lowering operators the number operators \eqref{number operators} have vacuum quantum numbers, so there are no zero- or four-dagger terms which may be dropped from the Wick contractions.
We are stuck computing four correlators,
\begin{align}
    C^{00}_{xy} &= \frac{1}{4}\left(
        C^{pp}_{xy} + C^{hh}_{xy} - C^{ph}_{xy} - C^{hp}_{xy}
        +
        \left\langle 1 -n^p_x - n^p_y+ n^h_x +n^h_y \right\rangle
        \right)  \nonumber\\
    C^{03}_{xy} &= \frac{1}{4}\left(
        C^{pp}_{xy} - C^{hh}_{xy} + C^{ph}_{xy} - C^{hp}_{xy}
        +
        \left\langle 1 - n^p_x - n^p_y + n^h_x - n^h_y\right\rangle
        \right)  \nonumber\\
    C^{30}_{xy} &= \frac{1}{4}\left(
        C^{pp}_{xy} - C^{hh}_{xy} - C^{ph}_{xy} + C^{hp}_{xy}
        +
        \left\langle 1 - n^p_x - n^p_y - n^h_x + n^h_y \right\rangle
        \right)  \nonumber\\
    C^{33}_{xy} &= \frac{1}{4}\left(
        C^{pp}_{xy} + C^{hh}_{xy} + C^{ph}_{xy} + C^{hp}_{xy}
        +
        \left\langle 1 - n^p_x - n^p_y - n^h_x - n^h_y \right\rangle
        \right)
    \nonumber \\
    \text{and we define }
    C^{\rho\rho}_{xy} &= C^{pp}_{xy} + C^{hh}_{xy} - C^{ph}_{xy} - C^{hp}_{xy}  \\
    C^{\rho n  }_{xy} &= C^{pp}_{xy} - C^{hh}_{xy} + C^{ph}_{xy} - C^{hp}_{xy}  \\
    C^{n   \rho}_{xy} &= C^{pp}_{xy} - C^{hh}_{xy} - C^{ph}_{xy} + C^{hp}_{xy}  \\
    C^{n   n   }_{xy} &= C^{pp}_{xy} + C^{hh}_{xy} + C^{ph}_{xy} + C^{hp}_{xy}
\end{align}
so that a $\rho$ superscript indicates the charge operator \eqref{spin bilinears}
and an $n$ superscript the total number operator.
The Wick contractions may be explicitly computed or built by rewriting the definition of the number operators \eqref{number operators} as
\begin{align}
    S^0_x &= \frac{1}{2}\left(1+n^h_x-n^p_x\right)
    &
    S^3_x &= \frac{1}{2}\left(1-n^h_x-n^p_x\right)
\end{align}
and using the particle-hole \eqref{particle-hole correlator} and particle-particle \eqref{particle-particle correlator} correlators and the one-point functions \eqref{one point functions}.
Note that $S^{1,2}$ cannot be correlated with $S^{0,3}$ because each term would not have the right constituent operator content to contract completely, so those correlators automatically vanish.

The doubly charged operators have simple Wick contractions.  Note that ${(S\pp)}^\dagger = {(S\mm)}$ so that
\begin{align}
    {(C\pp\mm)}_{xy}
    &=  \left\langle \adjoint{a}_x b_x \adjoint{b}_y a_y \right\rangle
    =   \left\langle \adjoint{a}_x a_y b_x \adjoint{b}_y \right\rangle
    =   \left\langle (\delta_{yx} - a_y \adjoint{a}_x) b_x \adjoint{b}_y \right\rangle
    =   \left\langle (\delta_{yx} - P_{yx}) H_{xy} \right\rangle
    \\
    C\mm\pp
    &=  \left[C{\pp\mm} \text{ with } P \leftrightarrow H\right].
\end{align}
Based on the exact results, a Lepage-style argument \cite{lepage1990proceedings} suggests these doubly-charged correlators should suffer from a signal-to-noise problem.


\subsection{Conserved Quantities}\label{sec:conserved quantities}

When the Hamiltonian takes a Hubbard-Coulomb-like form,
\begin{equation}
    H = -\sum_{xy} \left(
                            \adjoint{a}_x h_{xy} a_y
        +   \sigma_\kappa   \adjoint{b}_x h_{xy} b_y
        \right)
        + \frac{1}{2} \sum_{xy} \rho_x V_{xy} \rho_y
\end{equation}
some of the bilinears may correspond to conserved quantities.  For example, we can calculate the commutator with a local charge density operator,
\begin{align}
    [H, \rho_z]
        &= \left[\sum_{xy} \adjoint{a}_x h_{xy} a_y + \
        \sigma_\kappa \adjoint{b}_x h_{xy} b_y, \rho_z\right]   \\
        &= \sum_x - \adjoint{a}_x h_{xz} a_z + \sum_y \adjoint{a}_z h_{zy} a_y - \sigma_\kappa (a \leftrightarrow b)
\end{align}
where we immediately dropped the interaction term since the charge operator commutes with itself.
If we sum $z$ over all space the two terms cancel, so that the total charge
\begin{equation}
    \label{eq:total charge}
    Q = \sum_z \rho_z
\end{equation}
is conserved.
One similarly finds the total spins in each direction conserved,
\begin{align}
    \left[H, S^i \right] &= 0
    &
    S^i = \sum_z S^i_z
\end{align}
for $i\in\{1,2,3\}$ and in fact the total spin also commutes with the Hamiltonian
\begin{align}
    [H, S^2] &=0
    &
    S^2 &= \sum_i (S^{i})^2.
\end{align}
When the operators are conserved, their two-point correlation functions are constant,
\begin{align}
    C^{QQ}(\tau)
        &= \frac{1}{\mathcal{Z}}\tr{ Q(\tau) Q(0) e^{-\beta H}} 
        &&= \frac{1}{\mathcal{Z}}\tr{ e^{+H\tau}Q(0)e^{-H\tau} Q(0) e^{-\beta H}}
        \nonumber\\
        &= \frac{1}{\mathcal{Z}}\tr{Q(0)e^{-H\tau} Q(0) e^{-(\beta-\tau) H}} 
        &&= \frac{1}{\mathcal{Z}}\tr{Q(0) Q(0) e^{-\beta H}}
\end{align}
where we wrote the Heisenberg-picture $Q(\tau)$ in terms of the zero-time operator and the Hamiltonian and repeatedly used the fact that $Q$ commutes with the Hamiltonian.
We can turn this relation on its head and get an estimate for the equal-time correlator $\left\langle Q^2\right\rangle$ by averaging over the temporal separation,
\begin{equation}
     \left\langle Q^2 \right\rangle
     =
     \frac{1}{N_t} \sum_{\tau} C^{QQ}(\tau)
     =
     \frac{1}{N_t} \sum_{\tau} C^{\rho\rho}_{++}(\tau),
\end{equation}
where a $+$ subscript indicates that the spatial index is summed over---in this case, implementing \eqref{total charge}.
This same observation holds for the total spin operators $S^i$ (and therefore also for $S^\pm$), with the Hamiltonian shown above.
We can measure the mean-squared magnetization $\left\langle S^2 \right\rangle$ by
\begin{align}
    \left\langle S^2 \right\rangle
    =
    \sum_{i=1}^3
    \frac{1}{N_t} \sum_\tau
    C^{ii}_{++}(\tau)
\end{align}

On small test examples one observes numerically that the correlators are flat with the exponential discretization and seem linear with time in the diagonal discretization; averaging properly still yields good values---for an explicit example see the last appendix of \Ref{Wynen:2018ryx}.

\subsection{Numerical Results}\label{sec:bilinear results}

\begin{figure}[ht]
  \centering
  \includegraphics[width=0.77\textwidth]{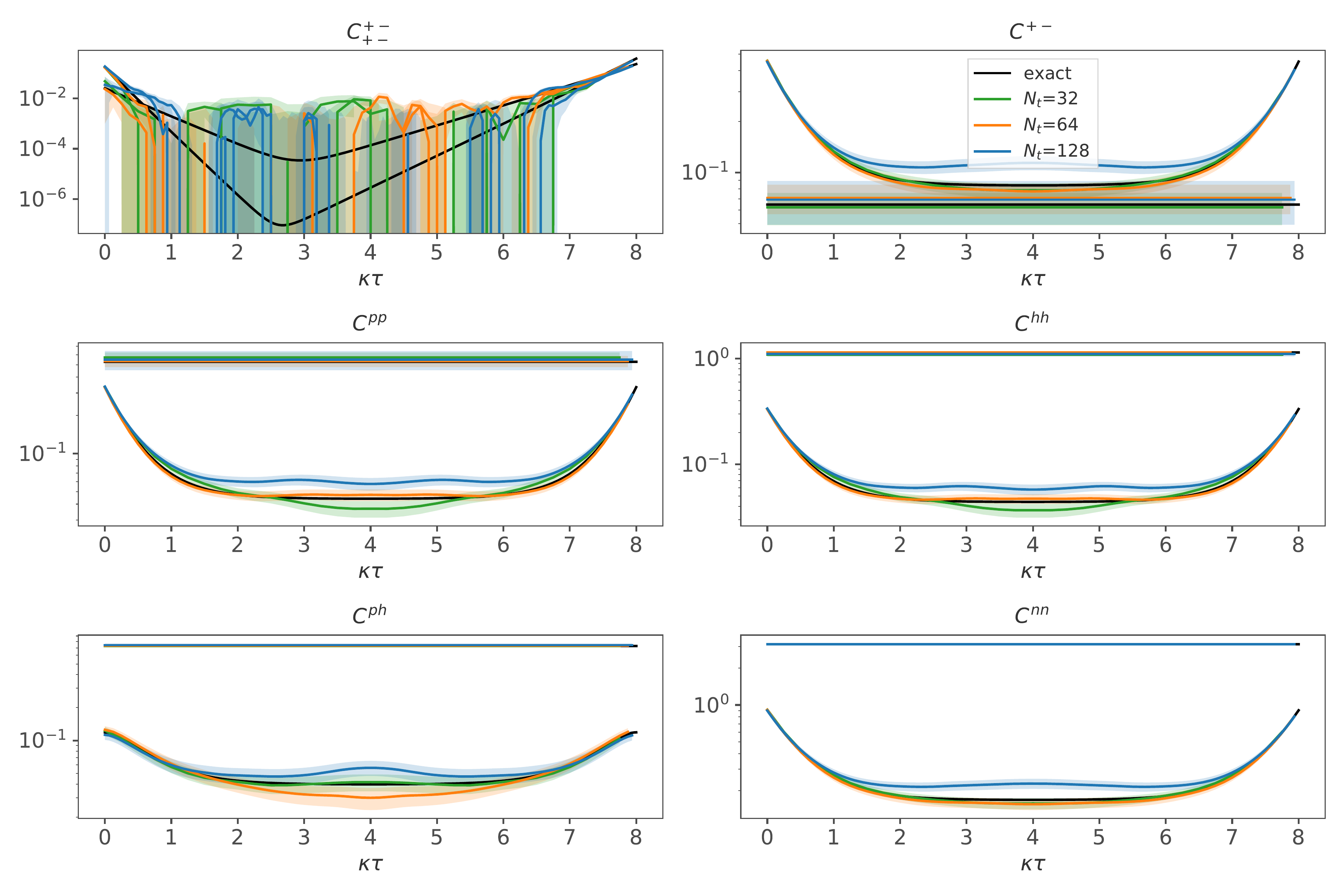}
  \caption{Numerical results for some correlation functions on a triangle lattice with $U/\kappa=3$, $\kappa\beta=8$ and $10^5$ configurations.\label{fig:triangle other corrs}}
\end{figure}

\begin{figure}[ht]
  \centering
  \includegraphics[width=0.77\textwidth]{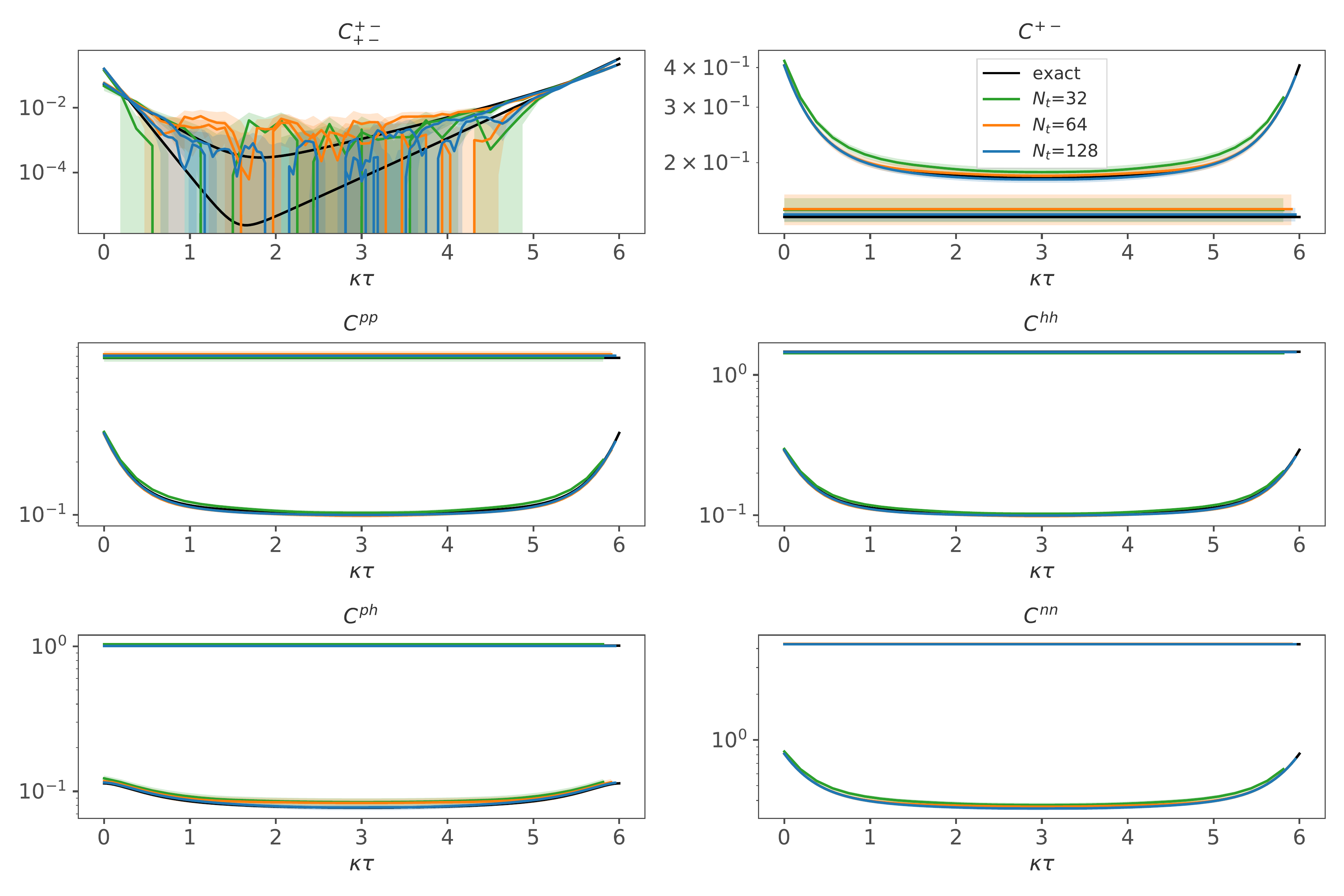}
  \caption{Numerical results for some correlation functions on a tetrahedron lattice with $U/\kappa=3$, $\kappa\beta=8$ and $10^5$ configurations.\label{fig:tetrahedron other corrs}}
\end{figure}

This section shows results for additional correlators that are not covered in the main text.
Correlators in Figures~\ref{fig:triangle other corrs} and~\ref{fig:tetrahedron other corrs} are computed on the same ensembles (``network'') as those in Sections~\ref{sec:triangle results} and~\ref{sec:tetrahedron results}, respectively.

\section{Coupling Layers}\label{sec:coupling layers}

As described above, the networks used here have a significant disadvantage, their Jacobians are expensive to compute.
Using coupling layers with a suitable coupling rule as introduced in \Ref{Dinh:2014nice} instead of dense layers, can effectively reduce that cost to zero in most applications.

For $x,y \in \mathds{R}^\Lambda$, let $\{A, B\}$ be a partition of the integer interval $\llbracket 1, \Lambda \rrbracket$. $\Lambda$ is the lattice spacetime volume $N_x\Nt$ and we restrict ourselves to even $\Lambda$ and $|A| = |B| = \Lambda/2$.
A coupling layer is defined as
\begin{align}
  f(x) =
  \begin{cases}
    y_A = x_A\\
    y_B = g(x_A, x_B)
  \end{cases}
\end{align}
Here, we focus on affine coupling layers which have a coupling rule of\,\footnote{The multiplicative part is typically written as $e^{s(x_A)}$ in the literature in order to simplify inversion. Since we do not need to invert the network here, we do not need the exponential.}
\begin{align}
  g(x_A, x_B) = x_B \odot s(x_A) + t(x_A)\ ,
\end{align}
where $\odot$ denotes element wise multiplication and $s$ and $t$ are arbitrary functions which can be parameterized through neural networks.
The Jacobian determinant of a network made out of layers, meaning $\NN(x) = f^n(f^{n-1}(\cdots f^1(x)))$, factorizes into separate determinants for each layer:
\begin{align}
  \det J = \det \pd{\NN}{x} = \det\bigg(\pd{f^{n}(x)}{x}\bigg) \det\bigg(\pd{f^{n-1}(x)}{x}\bigg) \cdots \det\bigg(\pd{f^{1}(x)}{x}\bigg)
\end{align}
In the case of affine coupling layers, the rows and columns of the layer Jacobians can be permuted under the determinants such that for each layer $A = \llbracket 1, \Lambda/2 \rrbracket$, $B = \llbracket \Lambda/2+1, \Lambda \rrbracket$ thus making the matrices triangular and the determinant fast to compute:
\begin{align}\label{eq:coupling layer jacobian}
  \det\bigg(\pd{f}{x}\bigg)
  = \det
  \begin{pmatrix}
    \mathds{1}_{\Lambda/2} & 0\\
    \pd{y_B}{x_A} & s(x_A)
  \end{pmatrix}
  = \prod_{i=1}^{\Lambda/2}\, {s(x_A)}_i\ ,
\end{align}
where $s(x_A)$ is to be understood as a diagonal matrix in the second expression.

This procedure can not be applied with the approach taken in this work as the transformation Eq.~\eqref{network transform} has Jacobian (see Eq.~\eqref{network jacobian})
\begin{align}
  \det J[\phi] = \det \bigg(\one + i\pd{\NN(\phi)}{\phi}\bigg)\ .
\end{align}
This determinant does not factorize and we can not bring the derivative of each layer into triangular shape.
$A$ and $B$ can also not be chosen such that the product of layers $\partial \NN(\phi) / \partial \phi$ itself is triangular because that would mean that not all components of $\phi$ could influence all components of $\tilde{\phi}$ which would limit the expressive power of the network.

The root cause for the problem is the different treatment of real and imaginary parts in Eq.~\eqref{network transform} because it leads to the ``$\one +$'' in the determinant.
A potential solution would be changing the transformation to $\tilde{\phi} = \NN(\phi)$.
However, this requires networks which deal with complex numbers and forgoes the potential benefits of Eq.~\eqref{network transform} described in the main text.
It is not trivial to formulate complex valued neural networks.
One difficulty comes from the loss functions which are typically non-analytic.
This problem can be avoided by using Wirtinger calculus to construct an optimization procedure~\cite{Brandwood:1983cgo,Kreutz:2009cgo}.
Another difficulty is finding complex activation functions~\cite{Scardapane:2018cvn}.
Here, the problem is complicated further because the chain rule of Wirtinger derivatives contains a sum:
\begin{align}
  z = f(y),\, y = g(x) \quad \Rightarrow \quad \pd{z}{x} = \pd{f}{y}\pd{g}{x} + \pd{f}{y^\ast}\pd{g^\ast}{x}
\end{align}
This would ultimately produce a sum in the determinant of the Jacobian, preventing factorization.
The efficient Jacobian of Eq.~\eqref{coupling layer jacobian} can thus only be realized with holomorphic neural networks, which implies holomorphic activation functions.
For this first stage of our work we relied on professionally-optimized, third-party machine learning libraries but support for complex-valued networks is currently incomplete.
We plan to develop an implementation with support for complex-valued coupling layers and pursue the approach described here.


\bibliography{cns}
\end{document}